\documentclass[12pt]{article}
\usepackage{setspace}
\usepackage[top=1in, bottom=1in, left=1in, right=1in]{geometry}
\RequirePackage{amsthm,amsmath,amsfonts,amssymb}
\RequirePackage[colorlinks,citecolor=blue,urlcolor=blue]{hyperref}
\RequirePackage{graphicx}
\usepackage{enumerate}
\usepackage{multirow}
\usepackage{booktabs}
\usepackage{longtable}
\usepackage{enumitem}
\usepackage{natbib}
\usepackage{bbm}
\usepackage{authblk}
\usepackage[titletoc,title]{appendix}
\usepackage{lscape}
\usepackage{xr}
\theoremstyle{plain}
\theoremstyle{remark}
\newtheorem*{remark}{Remark}
\newtheorem{theorem}{Theorem}[section]
\newlist{enumsteps}{enumerate}{2}
\setlist[enumsteps,1]{label=Step \arabic*., leftmargin=2\leftmargin,labelwidth=*}

\newlist{enumit}{enumerate}{2}
\setlist[enumit,1]{label=\arabic*., leftmargin=1.25\leftmargin,labelwidth=*}

\DeclareSymbolFont{AMSb}{U}{msb}{m}{n}

\def\Var{\mbox{Var}}

\DeclareMathSymbol{\E}{\mathbin}{AMSb}{"45}

\newcommand{\EE}[1]{\E \left [ #1 \right ]}

\newcommand{\var}[1]{\Var \! \left ( #1 \right )}
\newcommand{\varest}[1]{\widehat{\Var }\! \left ( #1 \right )}
\newcommand{\vark}[1]{\Var_k \! \left ( #1 \right )}
\newcommand{\varZ}[1]{\Var_{\mid \mathcal{Z}} \! \left ( #1 \right )}
\newcommand{\varkest}[1]{\widehat{\Var}_k \! \left ( #1 \right )}
\newcommand{\varkZ}[1]{\Var_{k \mid \mathcal{Z}} \! \left ( #1 \right )}
\newcommand{\varj}[1]{\Var_j \! \left ( #1 \right )}
\newcommand{\varjZ}[1]{\Var_{j  \mid \mathcal{Z}} \! \left ( #1 \right )}

\newcommand{\ycondk}[1]{\bar{Y}_{k\mid \mathcal{Z}}(#1)}

\newcommand{\ycond}[1]{\bar{Y}_{\mid \mathcal{Z}}(#1)}

\newcommand{\Nzk}{N_{zk}}
\newcommand{\Nonek}{N_{1k}}
\newcommand{\Nzerok}{N_{0k}}
\newcommand{\Nzj}{N_{zj}}

\newcommand{\NzkZ}{N_{zk \mid\mathcal{Z}}}
\newcommand{\NonekZ}{N_{1k \mid\mathcal{Z}}}
\newcommand{\NzerokZ}{N_{0k \mid\mathcal{Z}}}
\newcommand{\NzjZ}{N_{zj \mid\mathcal{Z}}}
\newcommand{\NzlZ}{N_{z \ell \mid\mathcal{Z}}}

\newcommand{\nzk}{n_{zk}}

\newcommand{\nonek}{n_{1k}}
\newcommand{\nzerok}{n_{k0}}
\newcommand{\nzkZ}{n_{zk \mid\mathcal{Z}}}

\newcommand{\nonekZ}{n_{1k\mid\mathcal{Z}}}
\newcommand{\nzerokZ}{n_{0k\mid\mathcal{Z}}}

\newcommand{\yhat}[1]{\hat{Y}_\mathcal{S}(#1)}
\newcommand{\yhatk}[1]{\hat{Y}_k(#1)}

\newcommand{\szkZ}{s^2_{z k \mid \mathcal{Z}}}
\newcommand{\SzkZ}{S^2_{z k \mid \mathcal{Z}}}

\newcommand{\ate}{ATE}
\newcommand{\ateest}{\widehat{ATE}}
\newcommand{\atesest}{\widehat{ATE}_S}

\usepackage{color}

\definecolor{cmtcol}{rgb}{0.9, 0.36, 0}

\def\spacingset#1{\renewcommand{\baselinestretch}%
{#1}\small\normalsize} \spacingset{1}

 \providecommand{\tightlist}{%
  \setlength{\itemsep}{0pt}\setlength{\parskip}{0pt}}

\begin{document}

\title{\bf 
Stratified Sampling for Model-Assisted Estimation with Surrogate Outcomes}
\author[1]{Reagan Mozer\thanks{This work was supported by the Institute of Education Sciences, U.S. Department of Education, through Grant \textit{R305D220032}.}}
\author[2]{Nicole E. Pashley}
\author[3]{Luke Miratrix}

\affil[1]{Bentley University}
\affil[2]{Rutgers University}
\affil[3]{Harvard University}

\date{}
\maketitle

\begin{abstract}
In many randomized trials, outcomes such as essays or open-ended responses must be manually scored as a preliminary step to impact analysis, a process that is costly and limiting. Model-assisted estimation offers a way to combine surrogate outcomes generated by machine learning or large language models with a human-coded subset, yet typical implementations use simple random sampling and therefore overlook systematic variation in surrogate prediction error. We extend this framework by incorporating stratified sampling to more efficiently allocate human coding effort.
We derive the exact variance of the stratified model-assisted estimator, characterize conditions under which stratification improves precision, and identify a Neyman-type optimal allocation rule that oversamples strata with larger residual variance. 
We evaluate our methods through a comprehensive simulation study to assess finite-sample performance.
Overall, we find stratification consistently improves efficiency when surrogate prediction errors exhibit structured bias or heteroskedasticity. 
We also present two empirical applications, one using data from an education RCT and one using a large observational corpus, to illustrate how these methods can be implemented in practice using ChatGPT-generated surrogate outcomes. Overall, this framework provides a practical design-based approach for leveraging surrogate outcomes and strategically allocating human coding effort to obtain unbiased estimates with greater efficiency. 
While motivated by text-as-data applications, the methodology applies broadly to any setting where outcome measurement is costly or prohibitive, and can be applied to comparisons across groups or estimating the mean of a single group.
\end{abstract}

\noindent%
{\it Keywords:} text analysis, causal inference, stratified sampling, large language models
\vfill

\newpage
\spacingset{1.5}

\section{Introduction}
\label{sec:intro}

Randomized trials that use text data as outcomes often face a bottleneck: every document (essay, survey response, etc.) must be hand-coded by humans to measure the outcome, which is time-consuming and expensive.
To reduce this burden, one might use a large language model (LLM) to code the documents instead, and then estimate impacts using these auto-coded outcomes, resulting in a radical reduction of work.
While appealing, such an approach faces serious validity concerns: is the machine correct?
To validate, we would want to at least audit a sample of the machine coded documents, and possibly adjust the impact estimate to account for any found biases in the machine coding.

Survey sampling provides the tools to do this.
As laid out in, e.g., \cite{mozer2024more} or \cite{egami2023using}, researchers can estimate impacts using the machine predicted outcomes with a bias-correction step that uses the differences between the human-coded and machine-predicted outcomes within the coded sample to ensure unbiasedness.
This approach effectively uses ``untapped'' observations (i.e., those documents not manually scored) as a supplementary resource to increase the precision of an estimated treatment impact for the hand-coded outcomes.

This problem is quite general.
Instead of text, one might want to code x-rays for a follow-up to a medical trial, or assess emotional tone in recorded speech.
In any of these settings, machine learning (ML) or generative AI could be used to label units to generate a numeric outcome suitable for impact analysis.
In each of these cases, if a researcher did so, they would also want to adjust final impact estimates to ensure the auto-coding is faithful to the targeted measure.
Even more broadly, if we are in a context where we have easy to measure surrogate outcomes for all the units in our trial, and the ability to obtain a gold-standard actual outcome for a subset of those units, we can use these survey-sampling tools to guarantee that we have an unbiased estimate of the average treatment effect, as described by the gold standard outcome, while still taking advantage of all the data, including those data where we do not have the gold standard outcome.

In the original approach, as described in \cite{mozer2024more}, the sample of texts selected for human coding is chosen by simple random sampling within each treatment group from the full experimental sample, which treats all documents as equally informative.
One might wonder if a more purposeful sampling, e.g., sampling documents that are guessed to be more difficult for the machine to code, could give further precision gains as compared to this baseline approach.
This question is the motivation of this work.

In this work we propose using \emph{stratified} sampling within each treatment arm to make the human-coded sample more informative.
The key insight is that by strategically selecting documents for human coding, we can improve the precision of an estimated treatment effect without increasing the human coding burden.
The approach is based on the idea that there may be different subgroups across a corpus of documents that are similar in terms of the errors generated by ML or LLM predictions.
By sampling appropriately across these strata, we can reduce the variance of our overall estimate as compared to simple random sampling.

In particular, we investigate the following research questions:

\begin{enumerate}
\item Under what conditions does stratified sampling provide efficiency gains beyond simple random sampling when using the model-assisted estimation framework?
\item Given a fixed human coding budget, what is the optimal allocation across a given set of strata to minimize the variance of an estimated effect?
\item How can researchers identify strata and implement the stratified sampling approach in practice?	
\end{enumerate}

To answer these questions we derive analytic formula for the true variance of the stratified approach, and for the difference in variance between stratification and simple random sampling.
These formulae allow for determining the strata after randomization; the stratification can even be based on post treatment features such as the surrogate outcomes themselves, allowing for great flexibility.
Our resulting formulae reveal what aspects of residual variation, and what aspects of how the stratification groups these residuals, results in the largest variance gain.
These formulae also reveal an optimization problem to determine, with a fixed coding budget, the number of documents to sample within strata.
All our findings also immediately extend to the simpler problem of estimating the mean of a population using a subset of coded documents.
To ease adoption, we provide the \texttt{stratsampling} R package with code and an accompanying vignette to illustrate how to use these methods in practice.

Our paper proceeds as follows: We define the problem (Section~\ref{sec:background}) and provide the approach along with theoretical results for the variance reduction of stratification in Section~\ref{sec:methods}.
We also give a plug-in estimator for the variance so the tool can be used in practice and a Neyman allocation approach to optimize coding budget across strata to maximize precision.
After presenting our theoretical results, we then examine to what degree these approaches can provide gains, and when they provide the largest gains, in a set of simulations (Section~\ref{sec:sims}) and two empirical applications (Section~\ref{sec:application}).
The first application is for estimating the average treatment effect in an RCT, the second is estimating a mean in an observational study.

\section{Background \& Problem Setup}
\label{sec:background}

Let $N$ be the total number of units with $N_z$ units randomly assigned to treatment $z \in \{0,1\}$ ($N = N_0 + N_1$), and $n_z$ the total number of those units sampled for human coding in treatment arm $z$.
We use the potential outcomes framework, and assume the Stable Unit Treatment Value Assumption  \citep[SUTVA,][]{rubin_1980}.
Under this model, for each unit we would observe some complex outcome, e.g., the full text of an essay, under treatment and also under control.
We call these complex outcomes the raw potential outcomes $R_i(0)$ and $R_i(1)$.
Each unit $i$ in the sample also has a summary potential outcome under treatment, $Y_i(1)$, and control, $Y_i(0)$.
These $Y_i(z)$ are the true quantitative outcomes capturing some measure of interest for our units; these could, for example, be the code a correct application of some scoring rubric that the treatment text (for $Y_i(1)$) or control text (for $Y_i(0)$) of the document would receive.
The goal is to learn about the average treatment effect (ATE) on the true quantitative outcomes in the full sample, 
\[\ate = \bar{Y}(1) - \bar{Y}(0) = \frac{1}{N}\sum_{i=1}^N(Y_i(1) - Y_i(0)).\]
Let $Z_i \in \{0,1\}$ be the treatment assignment for unit $i$ and $\mathcal{Z} = (Z_1,\dots,Z_N)$ be the treatment assignment vector across all units.
Treatment assignment determines the observed raw outcomes for each unit, $R_i = Z_iR_i(1) + (1-Z_i)R_i(0)$, which we observe for all units.

If we somehow were able to observe $Y_i = Z_iY_i(1) + (1-Z_i)Y_i(0)$ also for all units, we could estimate the ATE with a difference in means

\begin{equation}
\label{eq:ate_full_coding}	
\ateest_{full} = \frac{1}{N_1} \sum_{i:Z_i=1} Y_i(1) - \frac{1}{N_0} \sum_{i:Z_i = 0} Y_i(0)   \equiv \ycond{1} - \ycond{0}.
\end{equation}

We use the ``$\mid \mathcal{Z}$'' notation to denote something conditioned on the treatment assignment; in this case we are taking the mean of the potential outcomes in the treated and control groups.
The well-known Neyman variance of our full estimator is 
\[ \var{ \ateest_{full} } = \frac{\var{Y_i(1)}}{N_1} + \frac{\var{Y_i(0)}}{N_0} - \frac{\var{\tau_i}}{N} ,
\]
where $\var{Y_i(z)}$ represents the variability of potential outcomes among all $N$ individuals under treatment $z$, and the last term is the variance of the individual treatment effects $\tau_i = Y_i(1) - Y_i(0)$.
The $\tau_i$ can never be observed due to the so-called fundamental problem of causal inference, and thus we usually drop the term and take the first two as a bound on the finite-sample variance.
See, e.g., \citet{ding2024first}.

Unfortunately, $\ateest_{full}$ applies only when we have $Y_i$ for all units.
In the context being examined here, we only observe the $Y_i$ for a subset of our units, by, e.g., human coding that subset.
Let indicator $S_i =1$ if unit $i$ was sampled for human coding (or for some other gold-standard, expensive coding) and $S_i = 0$ otherwise.
Then, for units with $S_i=1$, we observe $Y_i$.
For those units with $S_i=0$, we do not observe $Y_i$.

\subsection{Model-assisted estimation}

We assume that we have a pre-trained machine learner or other device $f(\cdot)$ that takes the raw unit data, $R_i$, and converts it to a predicted outcome, $\hat{Y}_i$.
We will call $f(\cdot)$ a \emph{predictor}.
This predictor could be, for example, an LLM asked to score a set of text documents.
Given $f(\cdot)$, each unit has an imputed outcome if assigned to treatment $z$, $\hat{Y}_i(z) = f( R_i(z) )$, and we denote by $\hat{Y}_i$ the observed imputed outcome that depends on the treatment unit $i$ was assigned to.

If units were sampled at random, the de-biased estimator suggested by \cite{mozer2024more} is

\begin{equation}
\label{eq:ma_srs}
\ateest = \hat{Y}(1) - \hat{Y}(0),
\end{equation}
where
\[\hat{Y}(z) =  \frac{1}{N_{z}}\sum_{i=1}^N 1_{Z_i=z} \hat{Y}_i + \frac{1}{n_z} \sum_{i=1}^N S_i 1_{Z_i=z}(Y_i - \hat{Y}_i),\]
and $1_{A}$ is the indicator function with $1_{A} = 1$ if the event $A$ in the subscript occurs and 0 otherwise.
Also see \citet{egami2023using} for a broader treatment of the problem that allows for covariate adjustment.

We let the potential residual from our predictor $f(\cdot)$ for unit $i$ under treatment $z$ be $e_i(z) =  Y_i(z) - \hat{Y}_i(z)$.
We only observe $e_i =  Y_i - \hat{Y}_i$ for units with $S_i=1$.

\citet{mozer2024more} shows that the above estimator can have improved precision vs. simply using the randomly sampled subset of units.
The precision gains depend on the predictive power of $f(\cdot)$ and, consequently, the general magnitude of the $e_i$'s.

\section{Using Stratified Sampling to Improve Precision}
\label{sec:methods}

The procedure presented in \citet{mozer2024more} operates under a simple random sample of the full set of units.
We ask if we can do better if we stratify the units first, and then sample subsets of each strata for human coding.

\subsection{The stratified estimator}\label{subsec:strat_est}

In our framing, the researchers first randomize the units to treatment and control.
They then observe, for all units, any baseline covariates, the directly observable post-treatment data (the $R_i$), and the predicted $\hat{Y}_i = f( R_i )$.
Next, based on all of this information, including the treatment assignment itself, the researchers would generate $K_{1|\mathcal{Z}}$ strata for the treated units and $K_{0|\mathcal{Z}}$ strata for the control.
Importantly, these strata can be build dynamically in response to all observable information up to this point, and also may depend on distribution of units in the treatment arms (hence the conditioning on $\mathcal{Z}$).
For example, the strata may be formed based on the observed distribution of a post-treatment measure within each treatment arm.
We cannot, however, have the target outcome $Y_i$; the strata therefore do not directly depend on those (although they can depend on the $\hat{Y}_i$ or any function of the $R_i$).

Under stratified sampling, once our units in treatment arm $z$ are divided into $K_{z\mid \mathcal{Z}}$ groups we select $\nzkZ$ units from group $k$'s $\NzkZ$ total units.
We note that the stratification could be done differently in each arm and use the notation $ \mid \mathcal{Z}$ to emphasize that the number of strata and membership of strata all can depend on which units ended up being assigned to treatment.
Let $m_{i \mid \mathcal{Z}} \in \{1,\dots, K_{Z_i\mid \mathcal{Z}}\}$ be the stratum that unit $i$ belongs to.

Given our stratified sample, we can extend the model assisted estimator to weight the residuals within strata based on the proportion of units sampled, as so:
\begin{equation}
\label{eq:ma_strat}
\atesest = \yhat{1} - \yhat{0},
\end{equation}
where each mean is a weighted average across the strata of that treatment arm:
\begin{equation*}
	\yhat{z} =  \frac{1}{N_{z}}\sum_{i=1}^N 1_{Z_i=z} \hat{Y}_i + \frac{1}{N_z} \sum_{k=1}^{K_{z\mid \mathcal{Z}}} \sum_{i: m_{i \mid \mathcal{Z}} = k}  \frac{\NzkZ}{\nzkZ} S_i1_{Z_i=z}(Y_i - \hat{Y}_i).
\end{equation*}

\subsubsection{The estimator's variance}

Our estimator, $\atesest$ (Equation~\ref{eq:ma_strat}), can have, similar to the unstratified sample, improved precision over just analyzing the human-coded subset, as shown in the following theorem:

\begin{theorem}\label{theorm:prop_est}
$\atesest$ is unbiased for $\ate$ and its variance is
\begin{align*}
&\var{\atesest} \\
&=  \EE{ \sum_{k=1}^{K_{1\mid \mathcal{Z}}} \frac{\NonekZ(\NonekZ-\nonekZ)}{N_1^2\nonekZ}\varkZ{ e_i(1) |\mathcal{Z}} +  \sum_{k=1}^{K_{0\mid \mathcal{Z}}}  \frac{\NzerokZ(\NzerokZ-\nzerokZ)}{N_0^2\nzerokZ}\varkZ{ e_i(0) |\mathcal{Z}}}\\
&\qquad +  \var{\ateest_{full}}
\end{align*}
where
\begin{align*}
\varkZ{ e_i(z) |\mathcal{Z}} &= \frac{1}{\NzkZ-1}\sum_{i:m_{i \mid \mathcal{Z}} = k}1_{Z_i=z}(e_i(z) - \bar{e}_{k\mid \mathcal{Z}}(z))^2,
\end{align*}
and
\[ \bar{e}_{k \mid \mathcal{Z}}(z) = \frac{1}{\NzkZ}\sum_{i:m_{i \mid \mathcal{Z}} = k}1_{Z_i=z}e_i(z) . \]
\end{theorem}


The proof of this theorem, and all subsequent ones, are in Supplementary Materials \ref{append:proofs}.
This proof is based on a classic variance decomposition:
\[\var{\atesest} = \EE{\var{\atesest|\mathcal{Z}}} + \var{\EE{\atesest|\mathcal{Z}}}.\]
The first term gives the first line and the second the last term in Theorem~\ref{theorm:prop_est}.
For the conditional expectation in the second term, we have
\[\EE{\atesest|\mathcal{Z}} = \EE{\ateest|\mathcal{Z}} = \ycond{1} - \ycond{0} = \ateest_{full}, \]
and so the second variance term in the decomposition is $\var{\ateest_{full}}$.

Theorem~\ref{theorm:prop_est} shows that the variance of our stratified estimator has a core variance term of the variance we would have gotten from a standard difference in means estimator if we could obtain all (treatment dependent) outcomes in the experiment.
In comparing the formula to the non-stratified formula in \citet{mozer2024more}, we see this very same final variance term.
This implies that stratified sampling can only improve precision if the rest of the variance, which involves the variability of the residuals within each stratum, is reduced.
In the next section, we present a variance difference formula to specifically articulate when stratification is most beneficial.

\begin{remark}
In the case that the strata do not depend on the treatment assignment, e.g., the strata are based on a pre-treatment covariate and the strata are determined before treatment assignment, then we can slightly simplify the variance expression to  
 \begin{align*}
&\var{\atesest}\\
 &=  \sum_{k=1}^K\left[\frac{1}{N_1^2} \EE{\frac{\Nonek( \Nonek - \nonek) }{ \nonek} }\vark{ e_i(1) } +  \frac{1}{N_0^2}\EE{\frac{\Nzerok( \Nzerok - \nzerok) }{ \nzerok}} \vark{ e_i(0) }\right]\\
&\qquad +   \var{\ateest_{full}}
\end{align*}
where
\[ \vark{ e_i(z) } = \frac{1}{N_{k}-1}\sum_{i:m_i = k}(e_i(z) - \bar{e}_{k}(z))^2.\]
That is, we can simplify the notation to drop the $\mathcal{Z}$ dependence and particularly simplify the expectation to note that the $\vark{}$ terms are now not dependent on $\mathcal{Z}$ and also independent of any strata sizes.

\end{remark}

\subsubsection{Estimating the variance}

A conservative plug-in estimator of $\var{\atesest}$ is
\begin{align*}
\varest{\atesest}  &=\sum_{k=1}^{K_{z\mid \mathcal{Z}}} \left[ \frac{1}{N_1^2}\frac{\NonekZ( \NonekZ - \nonekZ) }{ \nonekZ} \varkest{ e_i(1) } +  \frac{1}{N_0^2}\frac{\NzerokZ( \NzerokZ - \nzerokZ) }{ \nzerokZ} \varkest{ e_i(0) }\right]\\
&\qquad + \frac{\hat{S}^2(1)}{N_1} +  \frac{\hat{S}^2(0)}{N_0},
\end{align*}
where 
\begin{align*}
\varkest{ e_i(z) } &= \frac{1}{\nzkZ-1}\sum_{i:S_i = 1, Z_i =z, m_{i \mid \mathcal{Z}}=k}(e_i(z) - \widehat{\bar{e}}_{k \mid \mathcal{Z}}(z))^2
\end{align*}
with
\begin{align*}
\widehat{\bar{e}}_{k \mid \mathcal{Z}}(z) &= \frac{1}{\nzkZ}\sum_{i:S_i = 1, Z_i =z, m_{i \mid \mathcal{Z}}=k}e_i(z),
\end{align*}
and
\begin{align*}
\hat{S}^2(z) 
&= \frac{N_z}{N_z-1} \left[\sum_{k=1}^K \frac{\NzkZ}{N_z} \frac{1}{\nzkZ}\sum_{\substack{S_i = 1 \\ Z_i = z \\ m_{i \mid \mathcal{Z}} = k}} Y_i^2 - ( \bar{Y}^{obs}_{\mid \mathcal{Z}}(z))^2 + \sum_{k=1}^K\frac{\NzkZ^2}{N_z^2}\left(\frac{\NzkZ - \nzkZ}{\NzkZ }\right)\frac{\szkZ}{\nzkZ}\right]
\end{align*}
with
\[\bar{Y}^{obs}_{\mid \mathcal{Z}}(z) = \sum_{k=1}^K\frac{\NzkZ}{N_z}\bar{Y}^{obs}_{k\mid \mathcal{Z}}(z), \qquad \bar{Y}^{obs}_{k\mid \mathcal{Z}}(z) = \frac{1}{\nzkZ}\sum_{i:S_i=1, m_{i \mid \mathcal{Z}} =k, Z_i=z}Y_i(z),\]
\[\szkZ = \frac{1}{\nzkZ - 1} \sum_{\substack{S_i = 1 \\ Z_i = z \\ m_{i \mid \mathcal{Z}} = k}}(Y_i - \bar{Y}^{obs}_{k\mid \mathcal{Z}}(z))^2.\]

This plug in estimator is not friendly.
We have provided software to calculate it as part of \texttt{stratsampling}, our R package that supports this paper.


Referring back to the variance decomposition in Section~\ref{subsec:strat_est}, the components with $\varkest{ e_i(z) }$ estimate the expectation of the conditional variances, whereas the pieces with $\hat{S}^2(z)$ estimate the variance of the conditional expectation.
The $\hat{S}^2(z)$ in particular estimate the variance of potential outcomes across the full experimental population and are based on an estimator of population variance from a stratified sample from \cite{de2006sampling}.

The bias of this estimator (see Supplementary Materials~\ref{sm:var_est}) is
\[\E\left[\varest{\atesest}\right] - \var{\atesest} = \frac{\var{\tau_i}}{N}.\]
Thus, we have a conservative estimator of variance.
The bias is Neyman's correlation of potential outcomes problem; we would have it even for the oracle estimator with all units human coded.

\subsection{The variance reduction from stratification}

We next present the benefits of stratification, as compared to simple random sampling.
Our main result is the following theorem, which specifically captures the difference in uncertainty between the two options:

\begin{theorem}\label{theorm:var_reduction}

The difference (averaged across possible treatment assignments) in variance between the estimator that uses stratified sampling and the one that does not is
\begin{align*}
&\EE{\var{\ateest|\mathcal{Z}}} - \EE{\var{\atesest|\mathcal{Z}}} =  BS - WS
\end{align*}
where
\begin{align*}
 BS =&  \E\Bigg[ \sum_{k=1}^{K_{z \mid \mathcal{Z}}}\Bigg[\frac{N_{1} - n_{1}}{ N_{1} }\frac{\NonekZ}{n_1(N_1-1)}(\bar{e}_{k\mid  \mathcal{Z}}(1) - \bar{e}_{\mid \mathcal{Z}}(1))^2  + \frac{N_{0} - n_{0} }{ N_{0} }\frac{\NzerokZ}{n_0(N_0-1)}(\bar{e}_{k \mid \mathcal{Z}}(0) - \bar{e}_{\mid \mathcal{Z}}(0))^2\Bigg]\Bigg],\\
 \intertext{and}
  WS =&  \E\Bigg[ \sum_{k=1}^{K_{z \mid \mathcal{Z}}}\Bigg[ \left(\frac{ \NonekZ}{N_1^2}\frac{\NonekZ - \nonekZ }{ \nonekZ  } - \frac{\NonekZ-1}{N_1(N_1-1)}\frac{N_{1} - n_{1}}{n_1}\right) \varkZ{ e_i(1) |\mathcal{Z}}\\
  &\qquad+\left(\frac{\NzerokZ}{N_0^2} \frac{\NzerokZ - \nzerokZ }{\nzerokZ } - \frac{\NzerokZ-1}{N_0(N_0-1)}\frac{N_{0} - n_{0}}{n_0}\right)\varkZ{e_i(0)|\mathcal{Z}} \Bigg]\Bigg].
\end{align*}
\end{theorem}

$BS$ (between-strata variation) is a measure of how much the average residual (bias) varies between strata and $WS$ (within-strata variation) is a measure of the variability of residuals within each strata.
Large $BS$ means greater benefit of stratification.
Positive $WS$ means a reduction in the potential benefit of stratification.
Negative $WS$ would mean further benefit of stratification.

The $BS$ expression, which is always nonnegative, shows that stratification will generally be beneficial when the average residuals from the predictor are different between strata.
That is, we want to form strata where the predictions $\hat{Y}_i$ are likely to have the same sort (i.e., magnitude and direction) of error.
Stratification is usually successful when \textit{outcomes} are similar \citep[see, e.g.,][]{pashley2022block}; here, stratification in sampling will be useful if there is \textit{systematic bias} in the predictor that we can take advantage of.
It does not matter whether the initial outcomes are homogenous or not.

The $WS$ term can be positive or negative, 
To understand it more closely, consider it to be a weighted sum of the strata-level variance terms $\varkZ{e_i(z)|\mathcal{Z}}$, with weights
\begin{align*}	
 w_{zk} &= \frac{\Nzk}{N_z^2} \frac{\NzkZ - \nzkZ }{\nzkZ } - \frac{\NzkZ-1}{N_z(N_z-1)}\frac{N_{z} - n_{z}}{n_z} \\
 &\approx > \frac{\NzkZ}{N_z^2}\left( \frac{\NzkZ - \nzkZ }{\nzkZ } - \frac{N_{z} - n_{z}}{n_z} \right) \\
 &= \frac{\NzkZ}{N_z^2}\left( \frac{1}{p_{zk|\mathcal{Z}}} - \frac{1}{p_z}\right) ,
 \end{align*}
 where $p_{zk|\mathcal{Z}}$ is the proportion of units sampled in treatment arm $z$ of stratum $k$, and $p_z$ is the overall proportion of units sampled in treatment arm $z$.
 The symbols $\approx >$ in the second line is used to note that, while
 \[\frac{\NzkZ}{N_z} > \frac{\NzkZ-1}{N_z-1},  \]
these quantities should be fairly close in large samples.
The sign of these weights, and therefore the entire $WS$ term, depends primarily on the $p_{zk|\mathcal{Z}}$.
The within-strata variances themselves depend on how similar the residuals are within strata.
Thus, the $WS$ term will be small when units have similar biases in surrogate outcomes within strata.

First, under proportional sampling, where we pull the same fraction of units from each strata (with $p_{zk|\mathcal{Z}} \approx p_z$ for all $k$), we would expect the $w_{zk}$ to all be slightly positive but near zero. 
This would make the $WS$ term, overall, slightly positive but near zero as well.
(We cannot make these weights exactly zero, in general, given that we can only sample entire units and thus will end up sampling fractions close to, but not exactly equal to, the target.)

More broadly, if a weight $w_{zk}$ is negative, then it will contribute to making $WS$ smaller, potentially even negative.
We can make a given $w_{zk}$ negative by oversampling that stratum: note how as $p_{zk|\mathcal{Z}}$ grows, $1/p_{zk|\mathcal{Z}} - 1/p_z$ would go negative.

We are, unfortunately, faced with a constraint: the $p_{zk|\mathcal{Z}}$ depend on the $\nzkZ$ and we cannot grow all the $\nzkZ$ as their sum is fixed.
We would therefore want to, as much as possible, make the weights negative and large for large variance terms while not letting them grow too large and positive for the smaller variance terms.
Intuitively, this oversampling approach has appeal: sample more from the units that are harder to automatically score with the machine learner (the variance of the residuals reflects how difficult those units are to score).
In Section~\ref{sec:opt_allocation} we describe this intuition with an optimization problem, and we explore the benefits of this approach more in our simulation and empirical section. 
Unfortunately, the gains of oversampling high variance strata tend to be limited, as making some terms negative causes other terms to offset by being more positive.

Overall, if $WS$ tends to be small, the difference (and potential gains) of stratification are driven by $BS$, the contribution of variability between strata of residuals.
In other words, stratification has the most benefit when we are able to stratify in ways that isolate different directions of bias of the surrogate outcomes.

\subsection{Optimal allocation strategy}
\label{sec:opt_allocation}

Theorem~\ref{theorm:prop_est} depends on the number of units sampled within each stratum.
Given the formula it provides, we might ask how to reduce variance by allocating our total budget of $n$ units across the strata.

Consider choosing the number of units to sample for human coding from each stratum in treatment arm $z$ in order to minimize $\var{\yhat{z} \mid \mathcal{Z}}$.
Because sampling is done separately for each arm, we can optimize the allocation for each arm independently.
We can do the optimization dependent on the realized treatment $\mathcal{Z}$: If we find an optimal version of our stratified estimator, $\widetilde{\atesest}$, such that
\[\var{\widetilde{\atesest}|\mathcal{Z}} \leq \var{\atesest|\mathcal{Z}}\]
for all $\mathcal{Z}$ then we must also have
\[\EE{\var{\widetilde{\atesest}|\mathcal{Z}}}  \leq \EE{\var{\atesest|\mathcal{Z}}} .\]
The above captures the expected benefit of optimal allocation; we see this benefit in practice in our simulation study, below.

To achieve our optimal allocation, assume we have a coding budget of $n_z$ units for treatment arm $z$ and would like to find the optimal set of $\nzk$.
That is, we want
\[\arg\min_{\mathbf{n}_{z\mid \mathcal{Z}}=(n_{z1 \mid \mathcal{Z}},\dots,n_{zK \mid \mathcal{Z}})}\var{\yhat{z}|\mathcal{Z}} \text{ subject to } \sum_k\nzkZ = n_z.\]
The solution to this optimization is the Neyman Allocation \citep{Neyman34, robbins1952some}, which has been previously applied to the design-based potential outcome framework \citep[see, e.g.,][]{hahn2011adaptive, blackwell2022batch, ravichandran2024optimal}, making this a straightforward application, which we verify in Supplementary Materials~\ref{sm:opt_all}.
The solution is
\[\tilde{n}_{zk\mid \mathcal{Z}} = \frac{\NzkZ\sqrt{\varkZ{ e_i(z) |\mathcal{Z}}}}{\sum_{j=1}^{K_{z\mid \mathcal{Z}}}\NzjZ\sqrt{\varjZ{ e_i(z) |\mathcal{Z}}}}n_z.\]
In general, $\tilde{n}_{zk\mid \mathcal{Z}}$ will not be a whole number.
One can round to get an approximate solution, or use integer programming to find the true optima.

If using $\tilde{n}_{zk\mid \mathcal{Z}}$ is not feasible for all $k$ because $\tilde{n}_{zk\mid \mathcal{Z}} > \NzkZ$, then we have to adjust our optimization slightly.
Let $\mathcal{K}_s$ be the set of strata such that $\tilde{n}_{zk \mid \mathcal{Z}} > \NzkZ$.
Then instead of $\tilde{n}_{zk \mid \mathcal{Z}}$ use
\begin{equation}
\label{eq:optimal_alloc}	
n^\prime_{zk \mid \mathcal{Z}} = \begin{cases}
\NzkZ &\text{if } k \in \mathcal{K}_s\\
\frac{\NzkZ \sqrt{\varkZ{ e_i(z) |\mathcal{Z}}}}{\sum_{j \in\mathcal{K}_l}^{K_{z\mid \mathcal{Z}}}\NzjZ(\mathcal{Z}\sqrt{\varj{ e_i(z) |\mathcal{Z}}}}(n_z - \sum_{\ell \in \mathcal{K}_s}\NzlZ) &\text{if } k \notin \mathcal{K}_s.
\end{cases}
\end{equation}
If any $k$ is such that $n^\prime_{zk \mid \mathcal{Z}} > \NzkZ$, add $k$ to $\mathcal{K}_s$ and repeat this process until $n^\prime_{zk \mid \mathcal{Z}} \leq \NzkZ$ for all $k$.
See Supplementary Materials~\ref{sm:opt_all} for the derivation of this method.

\section{Simulation Study}
\label{sec:sims}

We conducted a comprehensive simulation study to evaluate the performance of the proposed stratified model-assisted estimators in a setting that mirrors practice: the estimand is the ATE on the human-coded outcome, and the surrogate (e.g., ML/LLM prediction) is an imperfect proxy with both systematic bias and random prediction error that can differ across strata.
We consider a finite population of size $N$, where only a subset of $n<N$ documents can be hand-coded due to cost constraints, while surrogate predictions are available for all $N$ units.
We vary the coding budget, the structure and magnitude of bias and residual variance across strata, the overall predictive strength of the surrogate outcome, and the strata configuration.
Our goal is to characterize \textit{when} stratified sampling helps relative to simple random sampling within the model-assisted estimation framework and \textit{by how much} it improves precision.

\subsection{The data generating process}

We first walk through the data generating process, and then state the factors we use in our simulation.
All code is available as part of our \texttt{stratsampling} package, making other simulation configurations easily explorable.
We begin by generating a finite population of size $N = 1000$, with units randomly assigned to treatment ($Z_i$ = 1) or control ($Z_i$ = 0) under a balanced completely randomized design (i.e., $N_0=N_1=500$).
Within each arm $z\in\{0,1\}$, we assign stratum membership $S_i\in\{1,\ldots,4\}$ according to arm-specific probabilities $\pi_{zk}$ determined by the simulation scenario (see ``Strata configuration'' in Table~\ref{tab:sim_factors}).

For each unit $i$, we then generate potential outcomes as
\[ Y_i(0)\sim \mathcal N(0,\sigma_Y^2),\qquad Y_i(1)=Y_i(0)+\tau.\]
Throughout, we set $\tau=0$ and $\sigma_Y=3$.
The ``true'' (i.e., human-coded) outcome observed for each unit is then $Y_i = (1-Z_i)Y_i(0) + Z_iY_i(1)$. 

For unit $i$ in stratum $k$, we generate a surrogate outcome $\hat{Y}_i$ representing the output of a hypothetical predictor applied to the data:
\begin{equation}
\label{eq:surrogate_dgp}
\hat{Y}_i = Y_i + b_k + \varepsilon_{ik},\qquad \varepsilon_{ik}\sim \mathcal N(0,\sigma^2_{\varepsilon,k}),
\end{equation}
where the $b_k$ are strata-specific bias terms that introduces systematic prediction errors within strata and the $\sigma^2_{\varepsilon,k}$ are within-strata residual variances.
Because strata are exogenous, the residual $e_i= Y_i-\widehat Y_i=-(b_k+\varepsilon_{ik})$ satisfies
\[
\mathbb E[e_i\mid S_i=k] = -\,b_k,\qquad \mathrm{Var}(e_i\mid S_i=k)=\sigma^2_{\varepsilon,k},
\]
which cleanly separates the \textit{between-strata} residual-mean dispersion (bias) and \textit{within-strata} residual variance levers that stratification and Neyman allocation exploit.
The overall variance of the noise of our surrogate outcome is 
\begin{equation}
Var( e ) = \sum_k w_{zk}\,(b_k-\bar b_z)^2 + \sum_k w_{zk} \sigma^2_{\varepsilon,k} = \mathrm{Var}(b)+\mathbb E[\sigma^2_\varepsilon] , \label{eq:resid_variance}	
\end{equation}
where $w_{zk}=N_{zk}/N_z$ are the arm-specific stratum weights.

In our simulation we control how much the $b_k$ and the $\sigma^2_{\varepsilon,k}$ vary across strata.
The more the $b_k$ vary, the more there are systematic differences in bias across strata.
The more the $\sigma^2_{\varepsilon,k}$ vary, the more there are differences in the surrogate's precision across strata.
To tune our simulation, first define a pseudo-$R^2$ measure within each treatment arm as $R^2 = 1-\var{e_i}/\var{Y_i}$; This $R^2$ is a measure of how good the surrogate is for predicting the gold standard outcome.

We set the $R^2$ to a desired value, and then set two additional quantities as two factors of the simulation: $v_k$, a preliminary set of scalings for $\sigma^2_{\varepsilon,k}$ (the relative heteroskedasticity across strata), and $b'_k$, a set of preliminary bias terms for the strata.
We will rescale these two sets of scalings to achieve our target $R^2$.
We could set all of this separately across treatment arms, but we keep the same values for each arm in the simulation.

To calculate the final $b_k$ and $\sigma^2_{\varepsilon,k}$ to hit our target $R^2$ within each treatment arm we do the following:
First, calculate a residual variance normalizing constant $V = \sum_k w_{zk}v_k = \mathbb E[\sigma^2_\varepsilon]$; we use this constant to keep the overall proportion of the variation in our residual noise due to bias the same regardless of which variance scaling we select.
Second, recenter the $b'$ to ensure zero overall population bias (i.e., shift the $b'$ so $\sum_k w_k b'_k=0$ where $w_k$ is the proportion of units in stratum $k$).

Third, calculate a scaling constant $c$ as
\[
c = \frac{\sigma_Y^2(1-R^2)}{\mathrm{Var}(b')+\mathbb E[v]/V}= \frac{\sigma_Y^2(1-R^2)}{\mathrm{Var}(b')+ 1}.
\]
The second equality is because of our normalizing constant $V$.
Our $c$ comes from wanting to set $Var(e)$ to $(1 - R^2) Var(Y_i)$, so $c$ divides this quantity by the initial variance (see Equation~\ref{eq:resid_variance}).

With our final $b_k = \sqrt{c} b'_k$ and $\sigma_{\varepsilon,k}^2=c v_k / Z$, we can generate our data as described above.
The $c$ controls the overall magnitude of the residual variance and bias values so our surrogates achieve the target pseudo-$R^2$ in the scale of our original outcome $Y_i$.
The $V$ decouples the variation in the $\sigma^2_\varepsilon$ with the overall bias-variance balance, which is in turn controlled by the variation in the $b_k$.
We note that, under this DGP, introducing greater bias by spreading apart the $b'$ will reduce the degree of heteroscedasticity of the variance due to the overall $R^2$ constraint.

\subsection{Simulation factors}
Our simulations are structured as a factorial experiment across the factors used in our described DGP.
We list the factors on Table~\ref{tab:sim_factors} and discuss them in the following:

\begin{enumerate}
\tightlist
	\item \textbf{Bias}: Bias is encoded via the $b'_k$ term, with the following values:
		\begin{itemize}
		\tightlist
		\item None: $b'_k=0$ for all $k$	
		\item Small: bias increases linearly across strata with $b_k=-0.25\sigma_Y, -0.08\sigma_Y, 0.08\sigma_Y, 0.25\sigma_Y$
		\item Moderate: Twice as much bias as Small with $b_k=-0.5\sigma_Y, -0.17\sigma_Y, 0.17\sigma_Y, 0.5\sigma_Y$.
		\item Large: Twice as much bias as Moderate, with $b_k=-\sigma_Y, -0.34\sigma_Y, 0.34\sigma_Y, \sigma_Y$.
		\item Extreme contrast: Large bias ($\pm \sigma_Y$) concentrated in two strata with $b=-\sigma_Y,0,0,\sigma_Y$.
		\end{itemize}
\item \textbf{Residual variance}: We manipulate prediction error heterogeneity across strata as:
	\begin{itemize}
	\tightlist
	\item Homogeneous: Constant residual variance across strata ($v_k=1$ for all $k$)
	\item Heterogeneous: Residual variance linearly increasing across strata with $v_k=0.25, 1.50, 2.75, 4.00$.
	\item Extreme contrast: Very low/high variance (0.1X/10X baseline) concentrated in two strata ($v_k= 0.1, 1.0, 1.0, 10.0$).
	\end{itemize}
	Because the variance multipliers are weight-normalized (i.e., $\EE{v}=1$) to preserve the target pseudo-$R^2$, the multipliers shift the \emph{allocation} of residual variance across strata while keeping the overall $\Var(e_i)$ constant.
\item \textbf{Predictiveness of the surrogates}: We examine two levels of overall predictive strength between $Y$ and  $\hat{Y}$ with target $R^2$ values of 0.4 and 0.85, representing moderate and strong relationships between machine predictions and true outcomes.
\item \textbf{Strata configuration}: For each unit, we assign stratum membership $m_{i\mid\mathcal{Z}}$ using one of three configurations: 
	\begin{itemize}
		\tightlist

	\item Balanced-exact: exactly equal stratum sizes within each arm (i.e., $N_{zk\mid\mathcal{Z}} = N_z/K_z$)
	\item Balanced-approx: approximately equal stratum sizes within each arm ($\pi_{zk}=1/K_z$ for all $k$)
	\item Unbalanced: intentionally uneven strata sizes with $\pi_{zk}\sim Unif[0.2,0.8]$	
	\end{itemize}
\item \textbf{Coding budget}: We examine a range of fixed coding budgets determined by the annotated fraction $h\in\{0.10,0.20,\ldots,0.90\}$ which gives $n_z=\lfloor h\,N_z\rfloor$. 

\end{enumerate}

\begin{table}[htbp]
\centering
\caption{Simulation Factors}
\label{tab:sim_factors}
\begin{tabular}{p{4cm}p{10cm}}
\toprule
\textbf{Factor/Description} & \textbf{Levels Tested} \\
\midrule
Population Size &  $N=1000$  \\
Number of Strata &  $K=4$ \\
Outcome SD &  $\sigma_Y=3$ \\
True ATE & $\tau=0$ \\
\addlinespace
Coding Budget & $h$=10\%, 20\%, 30\%, 40\%, 50\%, 60\%, 70\%, 80\%, 90\% \\\addlinespace
Strata Configuration & \texttt{Balanced-exact} ($n_k=N/K=250$ per stratum), \texttt{Balanced-approx} (approx. equal strata sizes), \texttt{Unbalanced} (varying strata sizes) \\
\addlinespace

Bias in ML predictions across strata & \texttt{None}, \texttt{Small} (up to $0.25\sigma_Y$), \texttt{Moderate} (up to $0.5\sigma_Y$), \texttt{Large} (up to $\sigma_Y$), \texttt{Extreme-contrast} (up to $\sigma_Y$ concentrated in 2 strata) \\
\addlinespace
Residual Variance of ML predictions across strata  & \texttt{Homogeneous} ($\sigma_{e}$=1), \texttt{Heterogeneous} (Linearly increasing $\sigma_e$ from 0.25X to 4X baseline), \texttt{Extreme-contrast} (Very low/high variance concentrated in 2 strata) \\
\addlinespace

\addlinespace
Target $R^2$ & $R^2=0.4, 0.85$ \\
\bottomrule
\end{tabular}
\end{table}

The full grid implied by the above spans all combinations of bias pattern (5) $\times$ residual variance (3) $\times$ predictiveness (2) $\times$ strata configuration (3) $\times$ coding budget (9), yielding a total of 810 scenarios.
For each scenario, we apply a variety of sampling and estimation strategies, as described next.

\subsection{Estimation Methods}
In each iteration of our simulation, for each simulation scenario, we first generate a synthetic data set using the procedure outlined above.
From the generated data, we then compute four estimators of the average treatment effect:

\begin{enumerate}
\item \textbf{Subset estimator}: The simple difference in means estimator using only a coded subset selected by SRS within each treatment arm.
\item \textbf{Model-assisted (SRS) estimator}: The model-assisted estimator under SRS, given by (\ref{eq:ma_srs}), applied to a sample drawn as above.
\item \textbf{Stratified model-assisted estimator (proportional allocation)}: The stratified model-assisted estimator given by (\ref{eq:ma_strat}) with strata sizes $n_{zk\mid \mathcal{Z}}$ chosen using proportional allocation.
\item \textbf{Stratified model-assisted estimator (optimal allocation)}: The stratified model-assisted estimator (\ref{eq:ma_strat}) with sampling based on the optimal (Neyman) allocation given by (\ref{eq:optimal_alloc}). We use the true (oracle) residual variances from full human coding to compute the allocation, representing an upper bound on the potential gains from optimal allocation.
\end{enumerate}


For comparison, we also compute the oracle estimator (\ref{eq:ate_full_coding}) obtained by coding all $N$ documents. This provides a baseline for assessing the relative efficiency of our estimators compared to full coding.

We repeat this procedure for $R=1000$ iterations for each scenario.
Note that our simulation operates under a super population framework, drawing a new finite population of $N=1000$ units in each iteration. This differs from the finite population framework assumed in Section~\ref{sec:methods}, but the theoretical results still apply within each iteration (i.e., averaging across iterations corresponds to taking expectations over the super population).

\subsection{Evaluation Metrics}
For each simulation scenario, we compute the Monte Carlo bias, empirical standard error (SE), and  mean squared error (MSE) of each estimator across replications. To assess the precision gains achieved using stratified sampling, we also examine two relative efficiency metrics:
\begin{itemize}
	\item Variance reduction vs. SRS: percent reduction in empirical variance relative to the model-assisted (SRS) estimator.
	\item Variance inflation vs. full coding: ratio of empirical variance to the variance of the oracle estimator.
\end{itemize}

\subsection{Results}
\label{sec:sim_results}

We present key findings from our simulation study below.
For clarity, we first aggregate results across bias and residual variance factors to compare configurations with any between-strata bias (small, moderate, large, or extreme-contrast) versus  no bias, and to distinguish between configurations with any residual variance heterogeneity (heterogeneous or extreme-contrast) versus homogeneous variance.
We average across the strata configuration factor as we did not find it to meaningfully impact our results.
Additional results, including figures disaggregated by all levels of bias and residual variance as well as diagnostics confirming that all estimators are approximately unbiased and achieve nominal coverage, are provided in Appendix ~\ref{app:sim_results} of the Supplement.

Figure~\ref{fig:sims_emp_se} shows the true standard errors across a range of fixed coding budgets faceted by bias (rows) and residual variance (columns) of the ML predictions, with separate panels for high ($R^2=0.85$) and moderate ($R^2=0.40$) predictive accuracy.
As expected, we see that the model-assisted estimators uniformly dominate the simple subset estimator across all settings and budgets. This aligns with the theoretical results presented in \citet{mozer2024more} showing that the model-assisted estimator under SRS will provide efficiency gains over the subset estimator as long as $R^2>0$.

\begin{figure}[h]
\begin{center}
\includegraphics[width=0.85\textwidth]{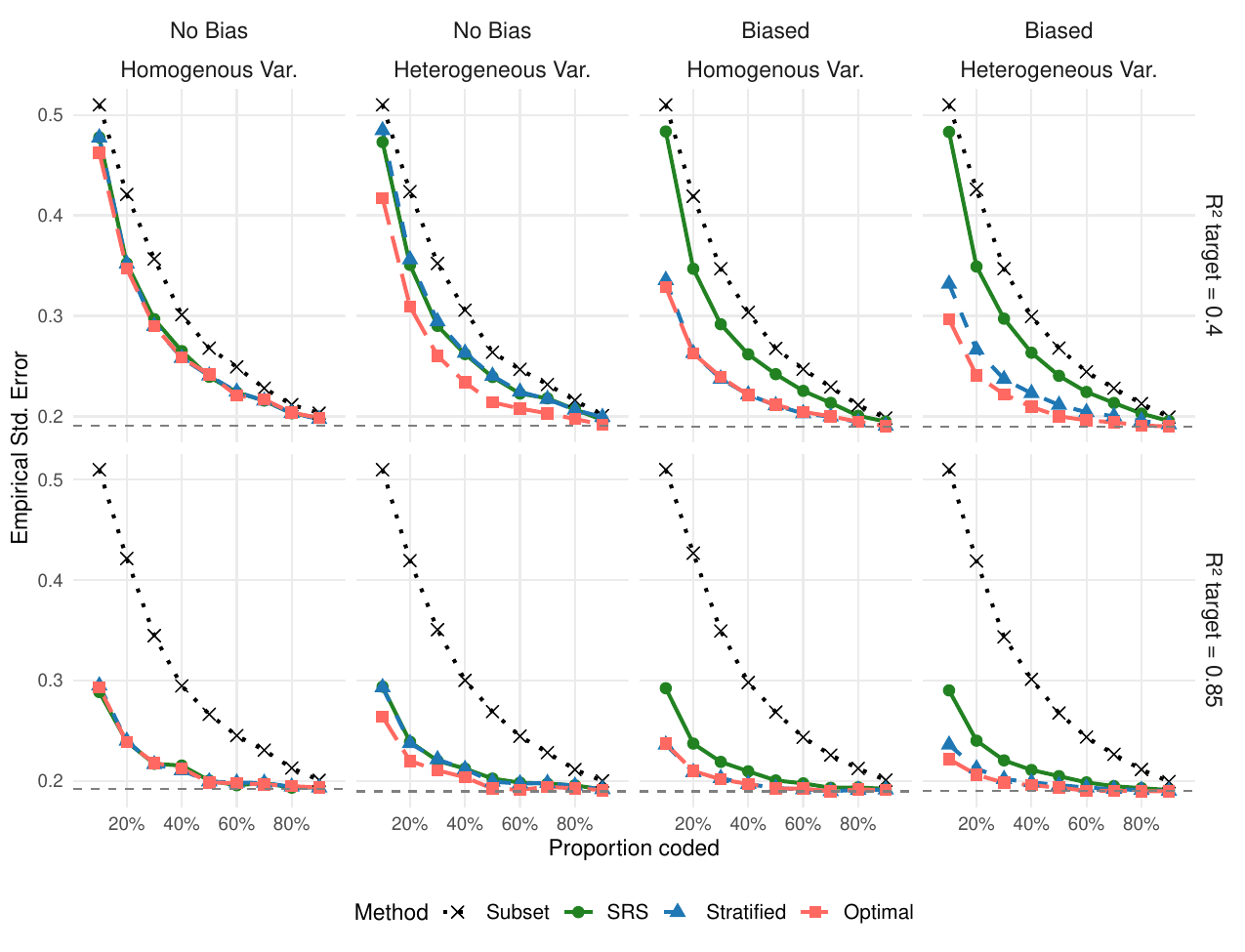}	
\end{center}
\caption{Average empirical SE of estimated treatment effect under different methods across simulation scenarios. Horizontal dashed line denotes standard error under full coding.}
\label{fig:sims_emp_se}

\end{figure}

\subsubsection{Benefits of stratified sampling}

Comparing the performance of the model-assisted estimator under SRS versus stratified sampling, there are two notable patterns.
First, stratified sampling improves precision when prediction error is structured (i.e.,  when the surrogate outcomes $\hat{Y}$ are biased and/or have heterogeneous residual variances across strata).
Conversely, when residuals are unbiased and have homogeneous variance across strata, the stratified estimator performs the same as the model-assisted estimator described in \citet{mozer2024more} under SRS. 
Second, stratified sampling with optimal allocation provides further efficiency gains compared to proportional allocation when residuals are heteroskedastic, particularly for low coding budgets and lower levels of predictive accuracy (i.e., when the surrogate outcomes are less predictive of the true human-coded outcomes).  

The largest efficiency gains from stratification arise when there is systematic bias in the surrogate outcomes.
When the residual means differ across strata, stratification effectively isolates these differences and ensures that the bias-correction step in the model-assisted estimator is estimated more precisely.
In these scenarios, stratified sampling substantially reduces variance relative to SRS, with gains increasing as the magnitude of between-strata bias grows.
See how on Figure~\ref{fig:sims_emp_se} the ``Stratified'' line is below the ``SRS'' line only under the biased conditions.

As a secondary concern, optimal stratification can further reduce the standard error when residual variances differ across strata (see the ``Optimal'' line in the two heterogeneous variance conditions of Figure~\ref{fig:sims_emp_se}).
This effect is especially pronounced at smaller coding budgets: note how the lines separate as we move to the left in the figures.

\subsubsection{Optimal allocation versus proportional allocation}
\begin{figure}[h]
\begin{center}
\includegraphics[width=0.85\textwidth]{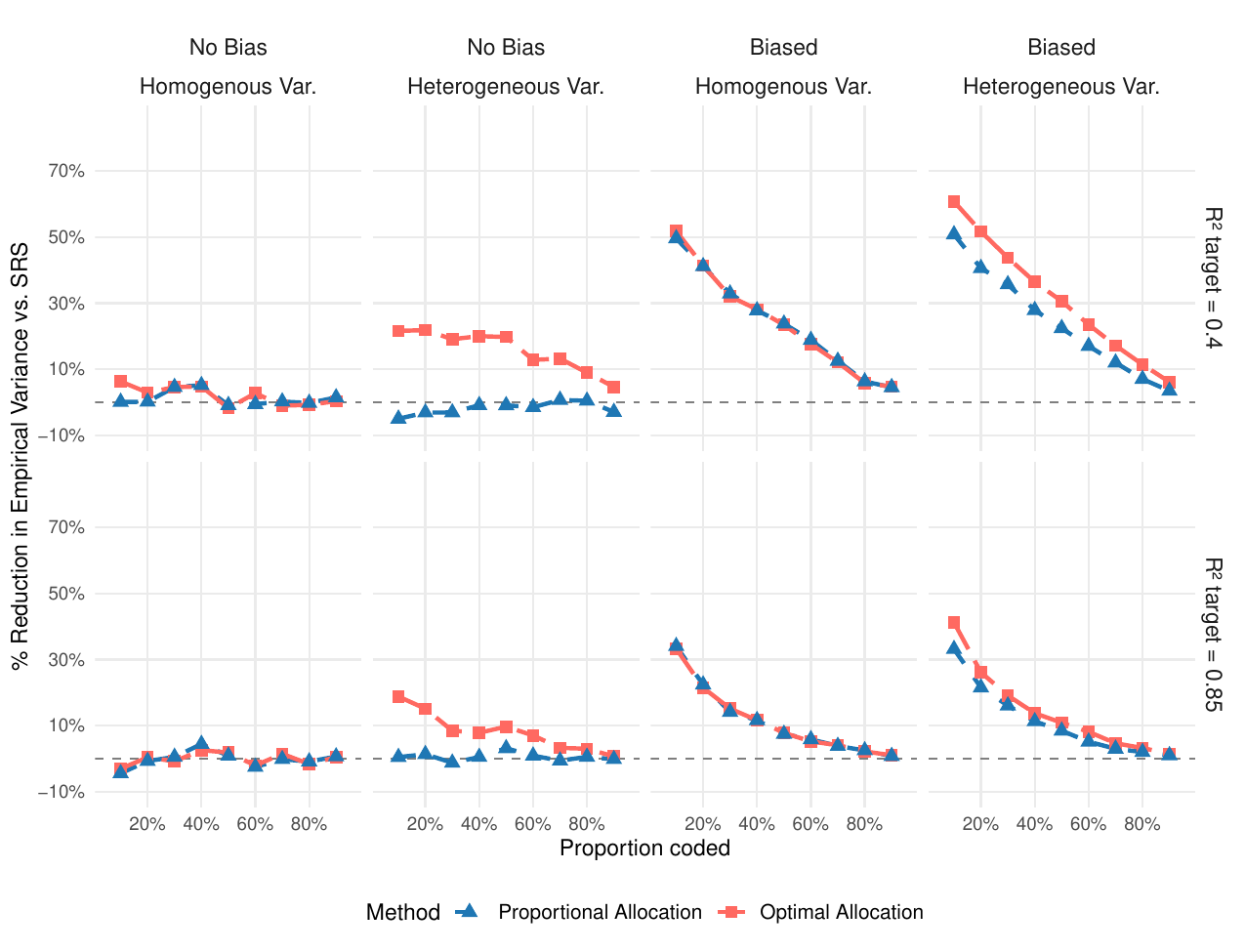}	
\end{center}
\caption{Percent reduction in empirical variance of the model-assisted estimator under stratified sampling compared to SRS.}
\label{fig:sims_re_vs_SRS}

\end{figure}

Figure~\ref{fig:sims_re_vs_SRS} shows the percent reduction in empirical variance achieved by stratified sampling relative to SRS, comparing proportional allocation to optimal (Neyman) allocation across simulation scenarios.

We see that stratification helps when there is bias (regardless of residual variance structure), optimal allocation helps when there is heterogeneity (regardless of bias), and using the stratified estimator with optimal allocation improves performance modestly over proportional allocation when residuals are both biased and heterogeneous.

Finally, the cost of using these tools when there is no bias or heteroskedasticity is apparently quite minimal, as shown by the pair of panels at the far left.

\subsubsection{Relative importance of simulation factors}

\begin{figure}[h]
\begin{center}
\includegraphics[width=0.8\textwidth]{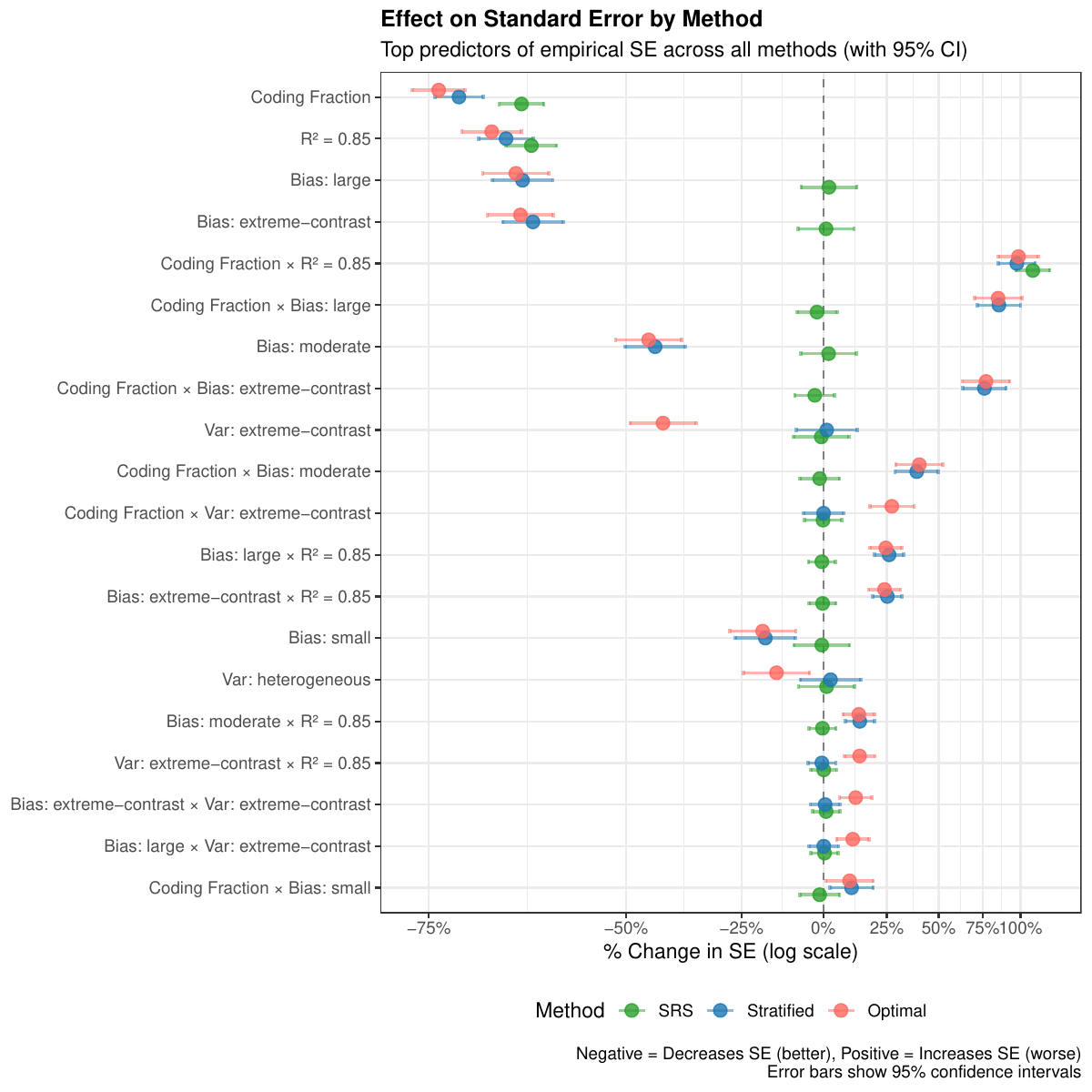}	
\end{center}
\caption{Relative influence of simulation design factors on empirical SE (log scale). Reference levels represent the configuration with no bias, homogeneous residual variance structure, balanced exact strata configuration, and a target $R^2$ of 0.4.}
\label{fig:sims_regression_coefs}
\end{figure}

To evaluate the relative importance of our different simulation design factors, we fitted regression models predicting the empirical standard errors (on the log-scale) for each of the model assisted estimators as a function of the coding budget, bias pattern, and residual variance structure, and target $R^2$ parameters, as well as all two-way interactions. Figure \ref{fig:sims_regression_coefs} shows the results as exponentiated coefficients, which represent the multiplicative change in standard error associated with each factor. Here, negative values indicate that the factor decreases the standard error (i.e., increases the efficiency) of the estimator relative to the baseline configuration (No bias, homogeneous residual variance, and a target $R^2$ of 0.4), and positive coefficients represent an increase in standard error relative to baseline.

A few patterns stand out. As expected, coding fraction has the largest effect on efficiency across all estimators. The effect of target $R^2$ is also substantial: higher predictive accuracy reduces standard errors for all model-assisted estimators, highlighting the importance of using good surrogate outcomes. 

More interestingly, the interaction between bias pattern and estimator type reveals where stratification helps most. For the model-assisted estimator using SRS, bias pattern has minimal effect on efficiency; this makes sense, as SRS treats all documents equally regardless of prediction error structure. In contrast, the stratified estimators show larger negative coefficients when residuals are biased across strata, indicating that these methods are effectively using stratification to achieve greater efficiency. In addition, the influence of residual variance structure on the performance of the stratified estimator  depends on allocation strategy: under optimal allocation, heterogeneous variance (particularly the extreme contrast configuration) is associated with lower standard errors, reflecting the benefits of oversampling from high-variance strata.

\section{Empirical Applications}
\label{sec:application}
In this section, we present two empirical applications to illustrate how the proposed stratified model-assisted framework can be implemented in practice.
The first application estimates a treatment effect from a randomized controlled trial, while the second estimates a population mean from a large observational corpus.
In both cases, we use ChatGPT to generate surrogate outcomes to demonstrate how researchers might use these methods in practice.

\subsection{MORE Study: Estimating a treatment effect}

Our first application uses data from a recent study by \citet{kim2021improving2}, who conducted a large-scale randomized controlled trial evaluating a content literacy intervention for early elementary students. 
In the study by \citet{kim2021improving2}, thousands of student essays were collected and later human-coded for writing quality, providing a rich setting in which ``gold-standard'' human-coded outcomes are available for all documents. 
Our goal in this section is to demonstrate how stratified sampling could have reduced the amount of human coding required to estimate the treatment effect of the MORE intervention on writing quality without sacrificing much precision.

Here, we first generated surrogate outcomes for all $N = 5,294$ essays using ChatGPT, prompted to score essays using the same rubric as the human coders.
In addition to providing a holistic quality score, we asked the model to report a confidence rating from 0--100 indicating its certainty in the assigned score.
Full prompting details are provided in Appendix~\ref{app:chatgpt}.
These LLM predictions serve as the surrogate outcome that a researcher would have at design time, before deciding which essays to human-code.

The correlation between the ChatGPT predictions and human scores was $\rho = 0.61$ overall, with similar predictive accuracy in both treatment arms ($\rho = 0.58$ in control, $\rho = 0.62$ in treatment).
This level of predictive power is typical of what might be achieved with current LLM technology on subjective scoring tasks and provides a realistic setting for evaluating our methods.

To identify an effective stratification, we considered a range of candidate stratification variables including the surrogate predictions $\hat{Y}_i$, the LLM's reported confidence scores, and essay length (word count).
For each candidate, we created stratifications based on quantile cuts (tertiles, quartiles, and quintiles) as well as interactions between pairs of variables.
We then evaluated these candidates using metrics that can be computed before any human coding begins: for example, we use the variance of the stratum means $\hat{Y}_i$ as a proxy for the between-strata variance that drives efficiency gains (see Section~\ref{sec:methods}).
See Appendix~\ref{app:strat_techniques} for a complete description of the candidate stratifications and evaluation process.

Based on this evaluation, we selected quartiles of the LLM predictions within each treatment arm as the final stratification, yielding four strata per arm.
This stratification can be implemented in practice because it depends only on the surrogate predictions $\hat{Y}_i$, which are available before any human coding begins.
Table~\ref{tab:reads_strat_v2} shows the characteristics of the resulting strata.
Note the systematic negative residuals: ChatGPT was systematically too lenient in its grading. 

\begin{table}[!h]
\centering
\caption{\label{tab:reads_strat_v2}Characteristics of final stratification for the MORE study determined by LLM-generated quality score (quartiles).}
\centering
\begin{tabular}[t]{rrp{2.5cm}rccc}
\toprule
$Z$ & $k$ & Predicted Score & $N_{zk}$ & Mean $\hat{Y}$ & Mean Resid. & Var(Resid.)\\
\midrule
0 & 1 & Low (0-2) & 911 & 1.47 & -0.46 & 0.56\\
0 & 2 & Med-Low (3) & 909 & 2.51 & -0.72 & 0.62\\
0 & 3 & Med-High (4) & 512 & 3.34 & -1.12 & 0.62\\
0 & 4 & High (5-7) & 309 & 4.29 & -1.75 & 0.77\\
\addlinespace
1 & 1 & Low (0-2) & 719 & 1.48 & -0.36 & 0.61\\
1 & 2 & Med-Low (3) & 840 & 2.51 & -0.48 & 0.69\\
1 & 3 & Med-High (4) & 625 & 3.34 & -0.88 & 0.62\\
1 & 4 & High (5-7) & 469 & 4.34 & -1.51 & 0.66\\
\bottomrule
\end{tabular}
\end{table}


Below, we demonstrate what a researcher might have observed by applying each of our estimators.
In particular, we, for each estimator, draw a single sample of 30\% ($h=0.3$) of the units for human coding (using SRS, proportional stratified sampling, or stratified sampling with optimal allocation) and then applying the estimator.
For the optimal allocation method, we used the Neyman allocation formula given in Equation~\ref{eq:optimal_alloc}, which requires estimates of the within-stratum residual standard deviations $\sqrt{\varkZ{e_i(z)|\mathcal{Z}}}$.
Since this was a retrospective analysis, we computed these using the oracle estimates from full human coding to illustrate the maximum potential gains from an oracle optimal allocation.

Table~\ref{tab:reads_point_ests_v2} shows the estimated treatment effects, standard errors, and confidence intervals for each method, with the oracle estimate obtained from full human coding serving as a benchmark. 

\begin{table}[!h]
\centering
\caption{\label{tab:reads_point_ests_v2}Results for the MORE study estimating the impact of the MORE intervention on writing quality scores from a single iteration with coding budget of h=30\%. Variance inflation is relative to full human coding (oracle).}
\centering
\begin{tabular}[t]{lrrrcc}
\toprule
Method & $n$ Coded & Estimate & SE & CI & Var. Inflation\\
\midrule
Oracle & 5294 & 0.333 & 0.027 & (0.28, 0.39) & 1.00\\
Subset & 1588 & 0.316 & 0.050 & (0.22, 0.41) & 3.28\\
SRS & 1588 & 0.377 & 0.047 & (0.28, 0.47) & 2.91\\
Stratified & 1588 & 0.341 & 0.043 & (0.26, 0.42) & 2.44\\
Stratified-Opt & 1588 & 0.380 & 0.042 & (0.30, 0.46) & 2.35\\
\bottomrule
\end{tabular}
\end{table}

There are substantial efficiency gains from using the model-assisted approach with stratified sampling.
The simple subset estimator, which ignores the surrogate predictions entirely, has a variance inflation factor of 3.28 relative to full coding (i.e., coding only 30\% of documents more than triples the variance).
Using the model-assisted estimator under simple random sampling reduces this inflation to 2.91, consistent with the results of \citet{mozer2024more}.
Notably, the stratified model-assisted estimators achieve even greater efficiency: proportional and optimal allocation yield variance inflation factors of approximately 2.44 and 2.35, respectively, representing a 16-19\% reduction in variance compared to the SRS approach and a 25-28\% reduction compared to the subset estimator.
For this stratification for this dataset, we see only small gains from optimal allocation; the modest variation in the residual variances across strata means little opportunity for improvements by shifting the coding budget to some strata vs. others.

To assess the sensitivity of these findings to the particular sample drawn, we repeated the analysis across 20 independent random samples at the 30\% coding budget.
The patterns were consistent across samples: the stratified estimators consistently outperformed the SRS and subset approaches. See Appendix~\ref{app:empirical_sens} in the Supplement for the full results and further discussion.

\begin{figure}[h]
\begin{center}
\includegraphics[width=0.8\textwidth]{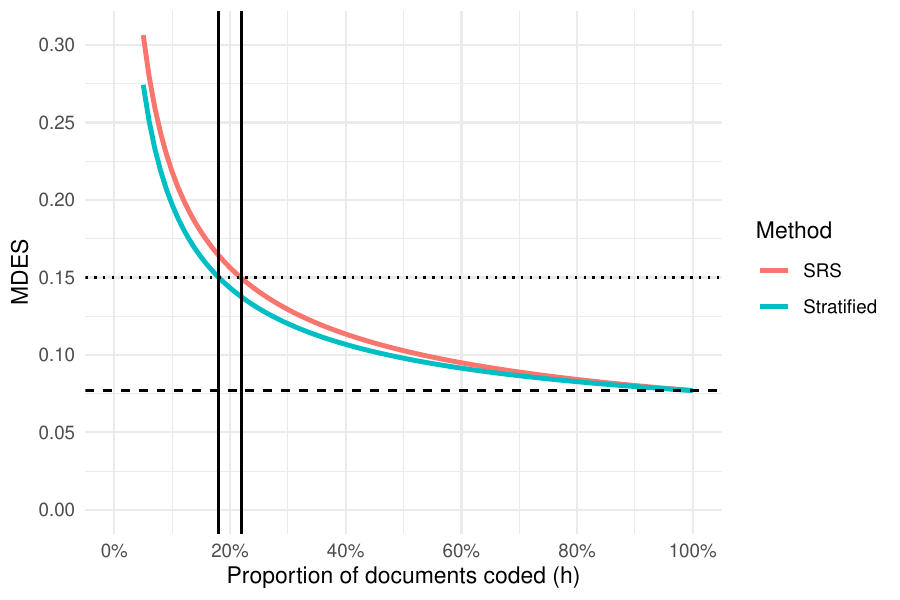}
\end{center}
\caption{MDES as a function of coding budget, using stratification or the SRS method.}
\label{fig:power_curve}
\end{figure}

Given pilot data, one can also use these formula to conduct a power analysis to determine what fraction of the corpus should be coded.
Figure~\ref{fig:power_curve} shows the Minimum Detectable Effect Size (MDES) as a function of how much of the corpus would be coded, using characteristics of the READS data for our design parameters.
We see with stratification, we can achieve an MDES of 0.15 with an 18\% coding effort vs. 22\%, or around a 20\% reduction.

\subsection{PERSUADE Corpus: Estimating a single mean}
Our second application illustrates how the proposed methods can be used to estimate a population mean in an observational setting. Here, we use the Persuasive Essays for Rating, Selecting, and Understanding Argumentative and Discourse Elements (PERSUADE) 2.0 corpus \citep{persuade2}, an open-source dataset that includes over 25,000 essays written by sixth- to twelfth-grade students in the United States along with human-coded holistic writing quality scores for each essay, which serve as our measured outcome $Y_i$.

As with the MORE application, we generated surrogate outcomes $\hat{Y}_i$ for all $N = 25,996$ essays using ChatGPT, prompted to provide holistic writing quality scores on the same scale as the human raters along with confidence ratings for each prediction (see Appendix~\ref{app:chatgpt} for further details).
The correlation between ChatGPT predictions and human scores was $\rho=0.67$, similar to the predictive accuracy observed in the MORE application.

We again considered multiple ways of constructing strata, with candidate stratification variables including the surrogate outcomes, the LLM's confidence scores, and essay length (word count).
We evaluated candidates using the variance of stratum means in $\hat{Y}_i$ as a proxy for potential efficiency gains, selecting the stratification that maximized this metric while maintaining adequate stratum sizes; see Appendix~\ref{app:strat_techniques} for additional details.

Using this approach, our final selected stratification uses the interaction between LLM predictions (binned into 4 groups) and essay length (below or above the median), yielding eight strata.
This choice was consistent with our hypothesis that: (1) very low or very high predicted scores might exhibit systematic bias, and (2) short essays (based on word count) may be more difficult for ChatGPT to predict than longer ones.
Table~\ref{tab:persuade_strat} displays the characteristics of the selected stratification.


\begin{table}[h]
\caption{Characteristics of final stratification for the PERSUADE corpus determined by predicted quality scores (4 groups) and essay length (word count; tertiles).}
\label{tab:persuade_strat}
\centering
\begin{tabular}[h]{lp{2.5cm}p{1.5cm}lcp{1.5cm}p{2cm}c}
\toprule
$k$ & Predicted Score & Word Count & $N_k$ & Mean $\hat{Y}$ & Avg. Length & Mean Resid. & Var(Resid.)\\
\midrule
1 & Low (1-2) & Low & 2672 & 2.64 & 236 & -0.95 & 0.29\\
2 & Low (1-2) & High & 592 & 2.63 & 693 & -0.55 & 0.64\\
\addlinespace
3 & Med-Low (3) & Low & 7991 & 3.99 & 268 & -1.75 & 0.33\\
4 & Med-Low (3) & High & 4854 & 3.99 & 516 & -0.93 & 0.53\\
\addlinespace
5 & Med-High (4) & Low & 2330 & 5.32 & 305 & -2.54 & 0.31\\
6 & Med-High (4) & High & 6309 & 5.32 & 572 & -1.56 & 0.47\\
\addlinespace
7 & High (5-6) & Low & 34 & 6.65 & 338 & -3.38 & 0.26\\
8 & High (5-6) & High & 1214 & 6.65 & 698 & -2.18 & 0.36\\
\bottomrule
\end{tabular}
\end{table}

Mean residuals range from $-0.55$ (stratum 2: low predictions, high word count) to $-3.38$ (stratum 7: high predictions, low word count).
The pattern indicates that the LLM tends to overestimate scores most severely for shorter essays with high or medium predicted scores.
Residual variances also vary considerably across strata, suggesting that optimal allocation could provide meaningful efficiency gains in this context.

Table \ref{tab:persuade_point_ests} presents results from a single sample using each estimation method at the 30\% coding budget ($n = 7,798$), with the oracle full-coding estimate serving as a benchmark.

\begin{table}[h]
\centering
\caption{Results from a single iteration with coding budget of h=30\%. Variance inflation is relative to full human coding (oracle).}
\label{tab:persuade_point_ests}
\centering
\begin{tabular}[t]{crcccc}
\toprule
Method & $n$ Coded & Estimate & \multicolumn{1}{c}{SE} & \multicolumn{1}{c}{95\% CI} & Var. Inflation \\ 
\midrule\addlinespace[2.5pt]
Oracle & 25996 & 2.857 & 0.006 & (2.844, 2.869) & 1.00\\
Subset & 7798 & 2.854 & 0.011 & (2.832, 2.876) & 3.31\\
SRS & 7798 & 2.853 & 0.010 & (2.834, 2.872) & 2.55\\
Stratified (Proportional) & 7798 & 2.854 & 0.009 & (2.837, 2.871) & 1.93\\
Stratified (Optimal) & 7798 & 2.852 & 0.009 & (2.835, 2.869) & 1.88\\
\bottomrule
\end{tabular}
\end{table}

Here, we again find that all methods produce estimates close to the oracle mean, with the model-assisted approaches providing substantially improved precision over the simple subset estimator.
The subset estimator, which ignores surrogate predictions, has a variance inflation factor of 3.31 relative to full coding.
Using the model-assisted estimator with SRS reduces this to 2.55, while stratified sampling with proportional allocation further reduces it to 1.93.
The optimal allocation achieves the best performance with a variance inflation factor of 1.88, representing a 26\% reduction in variance compared to SRS and a 43\% reduction compared to the subset approach.

We again see that optimal allocation provides only small gains compared to proportional allocation.
Despite moderate heterogeneity in the residual variances observed across strata in this application (see Table~\ref{tab:persuade_strat}), we again see that optimal allocation provides only small gains compared to proportional allocation. This is likely driven by large imbalance in stratum sizes relative to the total sample size; if strata with the largest residual variances are small relative to others, oversampling from these strata has little impact.

As with the MORE application, we conducted sensitivity analyses by repeating the estimation procedure across 20 independent samples at the 30\% coding budget.
Results were consistent across samples; the stratified estimators showed smaller variability in point estimates compared to alternatives, with optimal allocation providing modest gains over proportional allocation.
These results are presented in Appendix~\ref{app:empirical_sens} of the Supplement.

\subsection{Summary}

Both empirical applications demonstrate that stratified sampling provides meaningful efficiency gains over simple random sampling within the model-assisted estimation framework.
In the MORE study, stratification reduced variance by approximately 22\% relative to SRS when estimating a treatment effect; in the PERSUADE corpus, the reduction was approximately 21\% when estimating a population mean.
These gains arise because stratifying on the surrogate predictions (and, in the PERSUADE case, essay length) effectively isolates systematic patterns in surrogate prediction error.

Our results also show that, as expected, the relative benefit of optimal versus proportional allocation depends on the heterogeneity of residual variances across strata.
However, in both applications, optimal allocation provided only modest additional gains.

An important note, however, is that in both applications, we implemented optimal allocation using oracle information (i.e., we used the oracle residual variances to determine the Neyman allocation).
This approach represents a best-case scenario for optimal allocation, allowing us to isolate the potential gains from differential sampling across strata when the true residual structure is known.
In practice, researchers would need to estimate or guess the within-stratum residual variances prior to human coding in order to implement optimal allocation.
One approach for doing so would be to conduct a small pilot study by coding a small random sample (e.g., 5--10\% of units) to estimate stratum-level residual variances, then use these estimates to determine the optimal allocation for the remaining coding budget.
However, given that our results show only modest gains from optimal allocation even under ideal conditions, we recommend proportional allocation as the default approach for most applications.

\section{Discussion}
\label{sec:discuss}

Prior work has shown how auxiliary predictions, combined with a random sample of precisely coded units, can improve efficiency in randomized evaluations where outcomes are expensive to measure. We extend this work by examining whether stratified sampling can yield additional gains.

Our theoretical results show stratification helps primarily for strata where predictor bias is homogeneous within strata but varies across strata, and, as a secondary concern, where the predictor's accuracy varies across strata.
Empirically, we find that simply stratifying on the surrogate itself often achieves the first goal due to regression-to-the-mean effects: high predictions tend to be systematically too high, and low too low.
In both of our empirical applications, stratification on the surrogate (alone or in combination with document features) did notably improve efficiency, reducing variance by approximately 20\% relative to simple random sampling.
Optimal allocation, by contrast, provided little improvement over sampling proportional to strata size: it is hard to create strata where predictive accuracy varies substantially, and even when able to do so, the gains appear to be minimal.

Overall, our approach improves precision without changing estimands, introducing bias, or making strong modeling assumptions.
They apply both to two-group comparisons and to single-group designs, and could easily be extended to multiple group designs as well.
Efficiency gains do depend on the structure of prediction errors, and when errors are close to random, stratification yields little benefit.
Importantly, stratifying in such settings appears to also carry little cost.
Future work should extend these ideas to estimators that incorporate baseline covariate adjustment.
Regardless, we expect stratification to be useful in settings where rich covariates are available but outcomes are costly to obtain.
We hope the companion software\footnote{See the \texttt{stratsampling} R package and associated vignette.} makes using these tools accessible.

\subsection{Practical guidelines}
\label{subsec:guidelines}

We close with practical guidance for researchers seeking to implement these methods in their own applications.
The recommendations below are based on our theoretical results, simulation findings, and experience with the empirical applications.

\subsubsection{Analysis workflow}

A typical implementation of the stratified model-assisted estimation approach proceeds in five steps:

\begin{enumsteps}
\tightlist
\item \textbf{Generate surrogate outcomes.} Apply a predictor (e.g., an LLM prompted with the scoring rubric, or a trained ML model) to all $N$ units to obtain predicted outcomes $\hat{Y}_i$. The predictor should be independent of treatment assignment; if using a trained model, it should be fit on external data (e.g., from a pilot study) rather than human-coded outcomes from the current experiment.

\item \textbf{Construct strata.} Partition units into $K$ strata based on the surrogate predictions and/or other observable features. The primary goal is to make strata where errors tend to be homogeneous. This step uses only information available before human coding begins.

\item \textbf{Determine allocation.} Decide how many units to sample from each stratum ($n_k$), either using proportional allocation (sample sizes proportional to strata sizes) or optimal allocation (if estimates of within-stratum residual variances are available).

\item \textbf{Sample and code.} Draw a stratified random sample and obtain human-coded outcomes $Y_i$ for the sampled units.

\item \textbf{Estimate.} Compute the stratified model-assisted estimator and its variance estimate using the formulas in Section~\ref{sec:methods} or the accompanying software.
\end{enumsteps}

\subsubsection{Constructing a stratification}

\paragraph{Choosing stratification variables.}
The potential gains from stratified sampling compared to SRS depend directly on the choice of stratification. Based on our theoretical and empirical findings, we recommend stratifying on the surrogate outcomes as a primary approach.
Partitioning on quantiles of $\hat{Y}_i$ (e.g., quartiles or tertiles) is often effective because it captures systematic bias from regression-to-the-mean. If additional variables are available prior to human coding, incorporating secondary variables that are expected to correlate with prediction accuracy may provide additional benefits.
For text-as-data applications, promising candidates might include:
\begin{itemize}
\tightlist
    \item \textit{Document length}: Very short or very long documents may be more difficult to score automatically.
    \item \textit{Document type or source}: Different genres, prompts, or populations may have systematically different prediction error structures.
    \item \textit{Complexity indicators}: Readability scores, vocabulary diversity, or other structural features may predict where the model struggles.
\end{itemize}

When using multiple variables, we recommend creating strata using interactions between variables (e.g., tertiles of $\hat{Y}_i$ crossed with tertiles of word count). The \texttt{stratsampling} package provides functionality for constructing candidate stratification schemes in this manner.

\paragraph{Pre-coding diagnostics.}
Before finalizing the stratification, researchers can evaluate candidate stratifications using only the surrogate outcomes.
The \texttt{compare\_stratifications\_precoding()} function in our R package computes proxy metrics including the variance of stratum means (a proxy for $BS$) and the spread of within-stratum prediction variances.
Higher variance of stratum means suggests greater potential benefit from stratification.

\paragraph{Balanced versus unbalanced strata.}
Our simulations found that the strata configuration (balanced vs.\ unbalanced sizes) had minimal impact on estimator performance.
Using quantile-based cuts naturally produces approximately balanced strata, which simplifies implementation without sacrificing efficiency.

\subsubsection{Proportional versus optimal allocation}

As our simulation results and empirical applications suggest, the gains from optimal allocation over proportional allocation are generally modest (e.g., 5--10\% variance reduction in the best cases). 
In comparison, proportional allocation is simpler to implement, requires no pilot data, and performs nearly as well unless there is extreme heterogeneity in prediction error across strata.
In typical settings where estimates of within-stratum residual variance are not available prior to coding, proportional allocation serves as a reasonable default.

\subsubsection{Determining the coding budget}
Given some estimates for how predictive the surrogate is, and how well the strata could separate variances, we can use a simplification of our variance equation to estimate the final standard error of our process for any given coding fraction $h$.
Our software provides a method to plot power curves as a function of $h$ to assess the trade-off between coding less (smaller $h$) and larger standard errors (relative to coding everything).


\bibliographystyle{apalike}
\bibliography{refs.bib}

\clearpage


\begin{appendices}

\setcounter{page}{1}
\begin{center}
{\bf \Large  Supplementary Material\\for\\``Stratified Sampling for Model-Assisted Estimation with Surrogate Outcomes''}
\end{center}

\renewcommand{\theequation}{S\arabic{equation}}

Our supplementary materials to our main paper are organized as follows:
Section~\ref{append:proofs} provides proofs of the results in the main paper.
We then give additional simulation results.
These include the simulations broken out by scenario, rather than aggregated as they are in the main paper.
We finally provide further details on our two empirical applications, including the prompts used to generate the surrogate outcomes and details on how we stratify.

\section{Proofs}\label{append:proofs}

\subsection{Proof of Theorem \ref{theorm:prop_est}}

We have
\[\yhat{z} = \frac{1}{N_{z}(\mathcal{Z})} \sum_{i=1}^N 1_{Z_i=z}  \hat{Y}_i + \frac{1}{N_{z}} \sum_{k=1}^{K_{z\mid \mathcal{Z}}}\sum_{i: m_{i \mid \mathcal{Z}} = k} 1_{Z_i=z} S_i \frac{\NzkZ}{\nzkZ} \left( Y_i - \hat{Y}_i \right)\]
and can also define a with stratum version,
\[\yhatk{z} = \frac{1}{\NzkZ}\sum_{i: m_{i \mid \mathcal{Z}} = k} 1_{Z_i=z} \hat{Y}_i + \frac{1}{\nzkZ} \sum_{i: m_{i \mid \mathcal{Z}}=k} S_i 1_{Z_i=z}(Y_i - \hat{Y}_i).\]

For stratum $k$ we have
\begin{align*}
\EE{\yhatk{z}|\mathcal{Z}} &= \EE{\frac{1}{\NzkZ}\sum_{i: m_{i \mid \mathcal{Z}} = k} 1_{Z_i=z} \hat{Y}_i + \frac{1}{\nzkZ} \sum_{i: m_{i \mid \mathcal{Z}}=k} S_i 1_{Z_i=z}(Y_i - \hat{Y}_i)\Big|\mathcal{Z}}\\
&= \frac{1}{\NzkZ}\sum_{i: m_{i \mid \mathcal{Z}} = k} 1_{Z_i=z} \hat{Y}_i + \frac{1}{\nzkZ} \sum_{i: m_{i \mid \mathcal{Z}}=k} \EE{S_i\Big|\mathcal{Z}} 1_{Z_i=z}(Y_i - \hat{Y}_i)\\
&= \frac{1}{\NzkZ}\sum_{i: m_{i \mid \mathcal{Z}} = k} 1_{Z_i=z} \hat{Y}_i + \frac{1}{\nzkZ} \sum_{i: m_{i \mid \mathcal{Z}}=k}\frac{\nzkZ}{\NzkZ} 1_{Z_i=z}(Y_i - \hat{Y}_i)\\
&= \frac{1}{\NzkZ}\sum_{i: m_{i \mid \mathcal{Z}}=k} 1_{Z_i=z}Y_i \\
&  \equiv \ycondk{z}.
\end{align*}
So then
\[\EE{\yhat{z}|\mathcal{Z}} = \sum_k \frac{\NzkZ}{N_z}  \frac{1}{\NzkZ}\sum_{i: m_{i \mid \mathcal{Z}}=k} 1_{Z_i=z}Y_i  = \frac{1}{N_z}\sum_{i=1}^N 1_{Z_i=z}Y_i   \equiv \ycond{z}.\]
By standard randomization theory, $\EE{\EE{\yhat{z}|\mathcal{Z}}} = \bar{Y}(z)$, implying $\atesest = \yhat{1} - \yhat{0}$ is unbiased for $\ate$.

For the variance, we use a variance decomposition of
\[\var{\atesest} = \EE{\var{\atesest|\mathcal{Z}}} + \var{\EE{\atesest|\mathcal{Z}}}.\]

The conditional expectation is the same as the non-stratified case, leading to the variance of the conditional expectation being
\[ \var{\EE{\atesest|\mathcal{Z}}} =  \var{\ateest_{full}}. \]

Because $\mathcal{Z}$ is conditioned on, note that 
\[ \var{\yhatk{z}|\mathcal{Z}} =\var{ \frac{1}{\nzkZ} \sum_{i: m_{i \mid \mathcal{Z}}=k} S_i 1_{Z_i=z}(Y_i - \hat{Y}_i)\Bigg|\mathcal{Z}}=\var{ \frac{1}{\nzkZ} \sum_{i: m_{i \mid \mathcal{Z}}=k} S_i 1_{Z_i=z}e_i(z)\Bigg|\mathcal{Z}}.\]

We define
\begin{align*}
\varkZ{ e_i(z) |\mathcal{Z}} &= \frac{1}{\NzkZ-1}\sum_{i:m_{i \mid \mathcal{Z}} = k}1_{Z_i=z}(e_i(z) - \bar{e}_{k\mathcal{Z}}(z))^2\\
\end{align*}
where 
\[ \bar{e}_{k \mid \mathcal{Z}}(z) = \frac{1}{\NzkZ}\sum_{i:m_{i \mid \mathcal{Z}} = k}1_{Z_i=z}e_i(z) . \]
Then we have 
\[ \var{\yhatk{z}|\mathcal{Z}} =\frac{\NzkZ-\nzkZ}{\NzkZ}\frac{\varkZ{ e_i(z) |\mathcal{Z}}}{\nzkZ}.\]

We can use this to find the expectation of the conditional variance as
\begin{align*}
&\EE{\var{\atesest|\mathcal{Z}} }\\
&=  \EE{\sum_{k=1}^{K_{z\mid \mathcal{Z}}} \frac{\NonekZ^2}{N_1^2} \var{\hat{Y}_k(1)|\mathcal{Z}} +  \frac{\NzerokZ^2}{N_0^2}  \var{\hat{Y}_k(0)|\mathcal{Z}}}\\
&=  \EE{\sum_{k=1}^{K_{z\mid \mathcal{Z}}} \frac{\NonekZ^2}{N_1^2}\frac{\NonekZ-\nonekZ}{\NonekZ}\frac{\varkZ{ e_i(1) |\mathcal{Z}}}{\nonekZ} +  \frac{\NzerokZ^2}{N_0^2}  \frac{\NzerokZ-\nzerokZ}{\NzerokZ}\frac{\varkZ{ e_i(0) |\mathcal{Z}}}{\nzerokZ}}\\
&=  \EE{\sum_{k=1}^{K_{z\mid \mathcal{Z}}} \frac{\NonekZ(\NonekZ-\nonekZ)}{N_1^2\nonekZ}\varkZ{ e_i(1) |\mathcal{Z}} +    \frac{\NzerokZ(\NzerokZ-\nzerokZ)}{N_0^2\nzerokZ}\varkZ{ e_i(0) |\mathcal{Z}}}\\
\end{align*}

So overall, we have 
\begin{align*}
&\var{\atesest} \\
&=  \EE{\sum_{k=1}^{K_{z\mid \mathcal{Z}}} \frac{\NonekZ(\NonekZ-\nonekZ)}{N_1^2\nonekZ}\varkZ{ e_i(1) |\mathcal{Z}} +    \frac{\NzerokZ(\NzerokZ-\nzerokZ)}{N_0^2\nzerokZ}\varkZ{ e_i(0) |\mathcal{Z}}}\\
&\qquad +  \var{\EE{\ateest|\mathcal{Z}}}.  
\end{align*}

\subsection{Comparison of variance with and without stratification}
Because the conditional expectation of $\ateest$ and $\atesest$ are the same, we have
\[\var{\ateest} - \var{\atesest} = \EE{\var{\ateest|\mathcal{Z}}} - \EE{\var{\atesest|\mathcal{Z}}}\]

The conditional variance for the unstratified estimator can be written as
\begin{align*}
\var{\ateest|\mathcal{Z}} &= \frac{N_{1} - n_{1}}{ N_{1} }\frac{\var{ e_i(1) |\mathcal{Z}}}{n_1} + \frac{N_{0} - n_{0} }{ N_{0} }\frac{\var{ e_i(0)|\mathcal{Z}}}{n_0} 
\end{align*}
where
\begin{align*}
\var{ e_i(z) |\mathcal{Z}} &= \frac{1}{N_{z}-1}\sum_{i=1}^N1_{Z_i=z}(e_i(z) - \bar{e}_{\mid \mathcal{Z}}(z))^2\\
&= \frac{1}{N_z-1}\sum_{k=1}^{K_{z \mid \mathcal{Z}}}\left[(\NzkZ-1)\varkZ{e_i(z)|\mathcal{Z}} + \NzkZ(\bar{e}_{k \mid \mathcal{Z}}(z) - \bar{e}_{\mathcal{Z}}(z))^2 \right].
\end{align*}

First let's look at the difference in conditional variance between the unstratified and stratified estimator:
\begin{align*}
&\var{\ateest|\mathcal{Z}} - \var{\atesest|\mathcal{Z}}\\
  &= \frac{N_{1} - n_{1}}{ N_{1} }\frac{\var{ e_i(1) |\mathcal{Z}}}{n_1} + \frac{N_{0} - n_{0} }{ N_{0} }\frac{\var{ e_i(0)|\mathcal{Z}}}{n_0} \\
&\qquad -\sum_{k=1}^K \frac{\NonekZ^2}{N_1^2}\frac{ \NonekZ - \nonekZ }{ \NonekZ } \frac{ \vark{ e_i(1) |\mathcal{Z}} }{\nonekZ}- \frac{\NzerokZ^2}{N_0^2}  \frac{ \NzerokZ - \nzerokZ }{ \NzerokZ } \frac{ \varkZ{ e_i(0) |\mathcal{Z}} }{\nzerokZ}\\
 &= \sum_{k=1}^{K_{z \mid \mathcal{Z}}}\Bigg[\frac{N_{1} - n_{1}}{ N_{1} }\frac{\NonekZ-1}{n_1(N_1-1)}\varkZ{e_i(1)|\mathcal{Z}} + \frac{N_{0} - n_{0} }{ N_{0} }\frac{\NzerokZ-1}{n_0(N_0-1)}\varkZ{e_i(0)|\mathcal{Z}}\\
& \qquad +  \frac{N_{1} - n_{1}}{ N_{1} }\frac{\NonekZ}{n_1(N_1-1)}(\bar{e}_{k\mid  \mathcal{Z}}(1) - \bar{e}_{\mid \mathcal{Z}}(1))^2  + \frac{N_{0} - n_{0} }{ N_{0} }\frac{\NzerokZ}{n_0(N_0-1)}(\bar{e}_{k \mid \mathcal{Z}}(0) - \bar{e}_{\mid \mathcal{Z}}(0))^2 \\
&\qquad - \frac{\NonekZ^2}{N_1^2}\frac{ \NonekZ - \nonekZ }{ \NonekZ } \frac{ \varkZ{ e_i(1) |\mathcal{Z}} }{\nonek}- \frac{\NzerokZ^2}{N_0^2}  \frac{ \NzerokZ - \nzerokZ }{ \NzerokZ } \frac{ \varkZ{ e_i(0) |\mathcal{Z}} }{\nzerokZ}\Bigg]\\
\end{align*}

Then overall the difference is
\begin{align*}
&\EE{\var{\ateest|\mathcal{Z}}} - \EE{\var{\atesest|\mathcal{Z}}}\\
  &=  \E\Bigg[ \sum_{k=1}^{K_{z \mid \mathcal{Z}}}\Bigg[\frac{N_{1} - n_{1}}{ N_{1} }\frac{\NonekZ}{n_1(N_1-1)}(\bar{e}_{k\mid  \mathcal{Z}}(1) - \bar{e}_{\mid \mathcal{Z}}(1))^2  + \frac{N_{0} - n_{0} }{ N_{0} }\frac{\NzerokZ}{n_0(N_0-1)}(\bar{e}_{k \mid \mathcal{Z}}(0) - \bar{e}_{\mid \mathcal{Z}}(0))^2\\
  & \qquad  - \left(\frac{ \NonekZ(\NonekZ - \nonekZ) }{ \nonekZ N_1^2 } - \frac{(\NonekZ-1)(N_{1} - n_{1})}{n_1N_1(N_1-1)}\right) \varkZ{ e_i(1) |\mathcal{Z}}\\
  &\qquad- \left(\frac{\NzerokZ (\NzerokZ - \nzerokZ) }{N_0^2\nzerokZ} - \frac{(\NzerokZ-1)(N_{0} - n_{0})}{n_0N_0(N_0-1)}\right)\varkZ{e_i(0)|\mathcal{Z}} \Bigg]\Bigg]
\end{align*}

\subsection{Variance estimation}\label{sm:var_est}

Consider the variance of $\atesest$:
\begin{align*}
&\var{\atesest} \\
&=  \EE{\sum_{k=1}^{K_{z\mid \mathcal{Z}}} \frac{\NonekZ(\NonekZ-\nonekZ)}{N_1^2\nonekZ}\varkZ{ e_i(1) |\mathcal{Z}} +    \frac{\NzerokZ(\NzerokZ-\nzerokZ)}{N_0^2\nzerokZ}\varkZ{ e_i(0) |\mathcal{Z}}}\\
&\qquad +  \var{\widehat{ATE}_{full}}.
\end{align*}

The terms in the expectation of the first line can be unbiasedly estimated with a straightforward plug-in estimator:
\begin{align*}
\frac{1}{N_1^2}\frac{\NonekZ( \NonekZ - \nonekZ) }{ \nonekZ} \varkest{ e_i(1) } +  \frac{1}{N_0^2}\frac{\NzerokZ( \NzerokZ - \nzerokZ) }{ \nzerokZ} \varkest{ e_i(0) },
\end{align*}
where
\begin{align*}
\varkest{ e_i(z) } &= \frac{1}{\nzkZ-1}\sum_{i:S_i = 1, Z_i =z, m_{i \mid \mathcal{Z}}=k}(e_i(z) - \widehat{\bar{e}}_{k |\mathcal{Z}}(z))^2
\end{align*}
and
\begin{align*}
\widehat{\bar{e}}_{k |\mathcal{Z}}(z) = \frac{1}{\nzkZ}\sum_{i:S_i = 1, Z_i =z, m_{i \mid \mathcal{Z}}=k}e_i(z).
\end{align*}

For the second line, we have, from standard results,
\[ \var{\widehat{ATE}_{full}} = \frac{\var{Y_i(1)}}{N_1} + \frac{\var{Y_i(0)}}{N_0} - \frac{\var{\tau_i}}{N},\]
where
\[\var{Y_i(z)} = \frac{1}{N-1}\sum_{i = 1}^N (Y_i(z) - \bar{Y}(z))^2.\]
Because $\var{\tau_i}$ is not identifiable, we will focus on estimating $\var{Y_i(z)}$.

Define
\begin{align*}
&\varZ{Y_i(z)}\\
 &\equiv \frac{1}{N_z-1}\sum_{i: Z_i=z} (Y_i(z) - \ycond{z})^2.
\end{align*}

Because $\E[\varZ{Y_i(z)}] = \var{Y_i(z)}$, if we have an estimator that is conditionally (on $\mathcal{Z}$) unbiased for $\varZ{Y_i(z)}$, it will also be unconditionally unbiased for $\var{Y_i(z)}$.
Note the $\varZ{Y_i(z)}$ term is simply the usual $S^2(z)$ estimate of overall variance in outcomes found in the usual Neyman-style causal inference derivations of this nature (e.g., see \citet{miratrix2013adjusting}).

Consider a slight modification of the estimator from Equation 7.16 of \cite{de2006sampling}
\[\hat{S}^2(z) =\frac{N_z}{N_z-1} \left[\frac{1}{N_z}\sum_{k=1}^K\sum_{i: m_{i \mid \mathcal{Z}}=k, Z_i=z, S_i = 1}\frac{\NzkZ}{\nzkZ}Y_i^2 - ( \bar{Y}^{obs}_{\mid \mathcal{Z}}(z))^2 + \sum_{k=1}^K\frac{\NzkZ^2}{N_z^2}\left(\frac{\NzkZ - \nzkZ}{\NzkZ }\right)\frac{\szkZ}{\nzkZ}\right],\]
where
\[\bar{Y}^{obs}_{\mid \mathcal{Z}}(z) = \sum_{k=1}^K\frac{\NzkZ}{N_z}\bar{Y}^{obs}_{k\mid \mathcal{Z}}(z)\]
and
\[\szkZ = \frac{1}{\nzkZ - 1} \sum_{i:S_i = 1, Z_i =z, m_{i \mid \mathcal{Z}}=k}(Y_i - \bar{Y}^{obs}_{k\mid \mathcal{Z}}(z))^2\]
with
\[\bar{Y}^{obs}_{k\mid \mathcal{Z}}(z) = \frac{1}{\nzkZ}\sum_{i:S_i=1, m_{i \mid \mathcal{Z}} =k, Z_i=z}Y_i(z).\]

From standard stratified sampling results,
\begin{align*}
\Var{(\bar{Y}^{obs}_{\mid \mathcal{Z}}(z)\mid \mathcal{Z})} &=  \sum_{k=1}^K\frac{\NzkZ^2}{N_z^2}\frac{\NzkZ - \nzkZ}{\NzkZ }\frac{\SzkZ}{\nzkZ},
\end{align*}
where $\frac{\NzkZ - \nzkZ}{\NzkZ }$ is a finite population correction term,
and
\begin{align*}
\E[\szkZ | \mathcal{Z} ] = \SzkZ.
\end{align*}

We then have
\begin{align*}
&\E\left[\hat{S}^2(z)\mid \mathcal{Z}\right]\\
 &= \frac{N_z}{N_z-1} \left(\E\left[\frac{1}{N_z}\sum_{k=1}^K\sum_{i: m_{i \mid \mathcal{Z}}=k, Z_i=z, S_i=1}\frac{\NzkZ}{\nzkZ}Y_i^2 - ( \bar{Y}^{obs}_{\mid \mathcal{Z}}(z))^2 + \sum_{k=1}^K\frac{\NzkZ^2}{N_z^2}\left(1- \frac{\nzkZ}{\NzkZ}\right)\frac{\szkZ}{\nzkZ}\mid \mathcal{Z}\right]\right)\\
  &= \frac{N_z}{N_z-1} \Bigg(\frac{1}{N_z}\sum_{k=1}^K\sum_{i: m_{i \mid \mathcal{Z}}=k, Z_i=z}\frac{\NzkZ}{\nzkZ}\E\left[\mathbb{I}(S_i=1)\mid \mathcal{Z}\right]Y_i^2 - \left( \E\left[\bar{Y}^{obs}_{\mid \mathcal{Z}}(z)\right]^2 + \varZ{\bar{Y}^{obs}_{\mid \mathcal{Z}}(z)}\right)\\
  &\qquad \qquad \qquad + \sum_{k=1}^K\frac{\NzkZ^2}{N_z^2}\left( \frac{\NzkZ - \nzkZ}{\NzkZ}\right)\frac{\SzkZ}{\nzkZ}\Bigg)\\
  &= \frac{N_z}{N_z-1} \left[\frac{1}{N_z}\sum_{i: Z_i=z}Y_i^2 - \E\left[\bar{Y}^{obs}_{\mid \mathcal{Z}}(z)\right]^2 - \varZ{\bar{Y}^{obs}_{\mid \mathcal{Z}}(z)} + \varZ{\bar{Y}^{obs}_{\mid \mathcal{Z}}(z)}\right]\\
    &=\frac{N_z}{N_z-1} \left[ \frac{1}{N_z}\sum_{i: Z_i=z}Y_i^2 - \bar{Y}_{\mid \mathcal{Z}}(z)^2 \right]\\
    &= \frac{1}{N_z-1}\sum_{i: Z_i=z}(Y_i - \bar{Y}_{\mid \mathcal{Z}}(z))^2.
\end{align*}

Putting everything together, our overall estimator for variance is
\begin{align*}
\varest{\atesest} &=\sum_{k=1}^{K_{z\mid \mathcal{Z}}} \left[ \frac{1}{N_1^2}\frac{\NonekZ( \NonekZ - \nonekZ) }{ \nonekZ} \varkest{ e_i(1) } +  \frac{1}{N_0^2}\frac{\NzerokZ( \NzerokZ - \nzerokZ) }{ \nzerokZ} \varkest{ e_i(0) }\right]\\
&\qquad + \frac{\hat{S}^2(1)}{N_1} +  \frac{\hat{S}^2(0)}{N_0},
\end{align*}
which is conservative with bias $\frac{\var{\tau_i}}{N}$.
Interestingly, if the $n$s are much smaller than the $N$s, then the correlation of potential outcomes in the human coded sample is much reduced, and the bias will likely be small in terms of overall uncertainty.

\subsection{Proof of optimal allocation}
\label{sm:opt_all}

We will first show optimality of Neyman Allocation in this setting.
Recall that
\begin{align*}
\var{\yhat{z}|\mathcal{Z}} &= \sum_{k=1}^{K_{z\mid \mathcal{Z}}} \frac{\NzkZ^2}{N_z^2} \var{\yhatk{z}|\mathcal{Z}}\\
&= \sum_{k=1}^{K_{z\mid \mathcal{Z}}} \frac{\NzkZ^2}{N_z^2}  \frac{ \NzkZ - \nzkZ }{ \NzkZ } \frac{ \vark{ e_i(z) |\mathcal{Z}} }{\nzkZ},
\end{align*}
where $(\NzkZ - \nzkZ )/ \NzkZ$ is the finite population correction.

We can use the Lagrangian to find the $\nzkZ$ that minimize $\var{\yhat{z}|\mathcal{Z}}$.
To do so, we will introduce parameter $\lambda_0$ for the $\sum \nzkZ = n_z$ constraint and write our optimization as 
\[\min\left(\var{\yhat{z}|\mathcal{Z}} + \lambda_0\left[\sum_k\nzkZ - n_z\right]\right)\]
Let $\mathbf{n}_{z\mid \mathcal{Z}} = (n_{z1 \mid \mathcal{Z}}, \dots, n_{zK \mid \mathcal{Z}})$.

Define
\begin{align*}
f(\mathbf{n}_{z\mid \mathcal{Z}}) &= \var{\yhat{z}|\mathcal{Z}} + \lambda_0\left[\sum_k\nzkZ - n_z\right]\\
& = \sum_{k=1}^{K_{z\mid \mathcal{Z}}} \frac{\NzkZ^2}{N_z^2}  \frac{ \NzkZ - \nzkZ }{ \NzkZ } \frac{ \varkZ{ e_i(z) |\mathcal{Z}} }{\nzkZ}+ \lambda_0\left[\sum_k\nzkZ - n_z\right].
\end{align*}

Find minima using partial derivatives:
\begin{align*}
&\frac{\partial f(\mathbf{n}_{z\mid \mathcal{Z}})}{\partial \nzk} = - \frac{\Nzk^2}{N_z^2}  \frac{ 1 }{ \Nzk } \frac{ \vark{ e_i(z) |\mathcal{Z}} }{\nzk}  - \frac{\Nzk^2}{N_z^2}  \frac{ \Nzk - \nzk }{ \Nzk } \frac{ \vark{ e_i(z) |\mathcal{Z}} }{\nzk^2} + \lambda_0\\
&\frac{\partial f(\mathbf{n}_{z\mid \mathcal{Z}})}{\partial \lambda_0} = \sum_{k=1}^{K_{z\mid \mathcal{Z}}}\nzkZ - n_z.
\end{align*}
Now we set the partial derivatives equal to 0 and solve:
\begin{align*}
\nzkZ^2\lambda_0 &= \frac{\NzkZ^2}{N_z^2}\left[\frac{ \nzkZ }{ \NzkZ } +  \frac{ \NzkZ - \nzkZ }{ \NzkZ}\right]\varkZ{ e_i(z) |\mathcal{Z}}\\
&=\frac{\NzkZ^2}{N_z^2}\vark{ e_i(z) |\mathcal{Z}}\\
\sqrt{\lambda_0}\nzkZ&= \frac{\NzkZ}{N_z}\sqrt{\varkZ{ e_i(z) |\mathcal{Z}}}\\
\sqrt{\lambda_0}n_{z}&= \sum_{k=1}^{K_{z\mid \mathcal{Z}}}\frac{\NzkZ}{N_z}\sqrt{\varkZ{ e_i(z) |\mathcal{Z}}}\\
\implies & \nzkZ = \frac{\frac{\NzkZ}{N_z}\sqrt{\varkZ{ e_i(z) |\mathcal{Z}}}}{\sum_{j=1}^{K_{z\mid \mathcal{Z}}}\frac{\NzjZ}{N_z}\sqrt{\varj{ e_i(z) |\mathcal{Z}}}}n_z =  \frac{\NzkZ \sqrt{\varkZ{ e_i(z) |\mathcal{Z}}}}{\sum_{j=1}^{K_{z\mid \mathcal{Z}}}\NzjZ\sqrt{\varjZ{ e_i(z) |\mathcal{Z}}}}n_z
\end{align*}

For the case that the Neyman Allocation is infeasible, we can prove that the above is the optimal solution under the constraint that $\nzkZ \leq \NzkZ$.

We want
\[\arg\min_{n_{z1\mid \mathcal{Z}},\dots,n_{zK\mid \mathcal{Z}}}\var{\yhat{z}|\mathcal{Z}}\]
subject to $\sum_k\nzkZ = n_z$ and $\nzkZ \leq \NzkZ$.

We introduce slack variables $t_k^2$ ($\forall k$) and write our constraints as 
\begin{align*}
&\sum_k\nzkZ - n_z = 0\\
&\nzkZ - \NzkZ + t_k^2 = 0 \quad \forall \text{ } k
\end{align*}
We introduce parameters $\lambda_0$ and $\theta_k$ to write the function we will optimize, including the constraints, as
\[f(\mathbf{n}_z)  = \sum_{k=1}^{K_{z\mid \mathcal{Z}}} \frac{\NzkZ^2}{N_z^2}  \frac{ \NzkZ - \nzkZ }{ \NzkZ } \frac{ \varkZ{ e_i(z) |\mathcal{Z}} }{\nzkZ}+ \lambda_0\left(\sum_k\nzkZ - n_z\right) - \sum_{k=1}^{K_{z\mid \mathcal{Z}}}\theta_k(\nzkZ - \NzkZ + t_k^2).\]

Taking partial derivatives, we have
\begin{align*}
&\frac{\partial f(\mathbf{n}_{z\mid \mathcal{Z}})}{\partial \nzkZ} = - \frac{\NzkZ^2}{N_z^2}  \frac{ 1 }{ \NzkZ } \frac{ \varkZ{ e_i(z) |\mathcal{Z}} }{\nzkZ}  - \frac{\NzkZ^2}{N_z^2}  \frac{ \NzkZ - \nzkZ }{ \NzkZ } \frac{ \vark{ e_i(z) |\mathcal{Z}} }{\nzkZ^2} + \lambda_0 - \theta_k,\\
&\frac{\partial f(\mathbf{n}_{z\mid \mathcal{Z}})}{\partial \lambda_0} = \sum_{k=1}^{K_{z\mid \mathcal{Z}}}\nzkZ - n_z,\\
&\frac{\partial f(\mathbf{n}_{z\mid \mathcal{Z}})}{\partial \theta_k} = \NzkZ - \nzkZ - t_k^2.
\end{align*}

We find the minima by solving for the $\nzkZ$, setting these partial derivatives set to 0.
However, we now need to consider cases based on complimentary slackness conditions.
That is, for each $k$ either $\theta_k = 0$ and $t_k^2 >0$ OR $\theta_k \neq 0$ and $t_k^2 =0$.

\noindent \textbf{Case 1:} If $\theta_k = 0$ and $t_k^2 >0$ for all $k$, we get the same optimality solution as before. However, this result is only feasible if the resulting $\nzkZ \leq \NzkZ$ for all $k$ (which ensures that $t_k^2 >0$ as required).

\noindent \textbf{Case 2:} Let $\theta_k = 0$ and $t_k^2 >0$ for $k \in \mathcal{S}_t$ and $\theta_k \neq 0$ and $t_k^2 =0$ for $k \in \mathcal{S}_t^C$.
For all $k \in \mathcal{S}_t^C$,
\[\NzkZ = \nzkZ \]
and therefore
\[\sum_{k \in  \mathcal{S}_t}\nzkZ  = n_z - \sum_{k \in  \mathcal{S}_t^C}\NzkZ.\]

Further, for all $k \in  \mathcal{S}_t$,
\begin{align*}
\sqrt{\lambda_0}\nzkZ&= \frac{\NzkZ}{N_z}\sqrt{\varkZ{ e_i(z) |\mathcal{Z}}}\\
\text{implying }
\sqrt{\lambda_0}(n_{z}  - \sum_{k \in  \mathcal{S}_t^C}\NzkZ )&= \sum_{k \in  \mathcal{S}_t}\frac{\NzkZ}{N_z}\sqrt{\varkZ{ e_i(z) |\mathcal{Z}}}.
\end{align*}
Then we have, for all $k \in  \mathcal{S}_t$,
\begin{align*}
\sqrt{\lambda_0}&=\frac{ \sum_{k \in  \mathcal{S}_t}\frac{\NzkZ}{N_z}\sqrt{\varkZ{ e_i(z) |\mathcal{Z}}}}{n_{z}  - \sum_{k \in  \mathcal{S}_t^C}\NzkZ},\\
\nzkZ &= \frac{\NzkZ\sqrt{\varkZ{ e_i(z) |\mathcal{Z}}}}{\sum_{k \in  \mathcal{S}_t}\Nzj\sqrt{\varjZ{ e_i(z) |\mathcal{Z}}}}(n_z - \sum_{k \in  \mathcal{S}_t^C}\NzkZ).
\end{align*}
In order for this solution to be feasible, we require 
\[\frac{\NzkZ\sqrt{\varkZ{ e_i(z) |\mathcal{Z}}}}{\sum_{k \in  \mathcal{S}_t}\NzjZ\sqrt{\varjZ{ e_i(z) |\mathcal{Z}}}}(n_z - \sum_{k \in  \mathcal{S}_t^C}\NzkZ) < \NzkZ\]
for all $k \in  \mathcal{S}_t$.
This then ensures that for all $k \in  \mathcal{S}_t$,
\[t_k^2 = \NzkZ - \nzkZ - \NzkZ - \frac{\NzkZ \sqrt{\varkZ{ e_i(z) |\mathcal{Z}}}}{\sum_{k \in  \mathcal{S}_t}\NzjZ\sqrt{\varjZ{ e_i(z) |\mathcal{Z}}}}(n_z - \sum_{k \in  \mathcal{S}_t^C} \NzkZ)  > 0\]

It's clear that among all solutions under Case 2 that are feasible, the smallest variance must be obtained with the minimal constraints on the sample size, implying we only set $\nzkZ = \NzkZ$ for those strata where it is strictly necessary.

\section{Additional simulation results}
\label{app:sim_results}
Here we present additional results from the simulation study described in Section~\ref{sec:sims}. While the main text aggregates results across bias patterns (no bias vs.\ any bias) and residual variance structures (homogeneous vs.\ any heterogeneity), here we show the full disaggregated results broken down by specific configurations of bias pattern, residual variance structure, and target $R^2$. The results show that the patterns observed in the main text hold consistently across all simulation configurations.

\subsection{Efficiency frontiers by configuration}

\begin{figure}[h]
    \centering
    \includegraphics[width=0.9\textwidth]{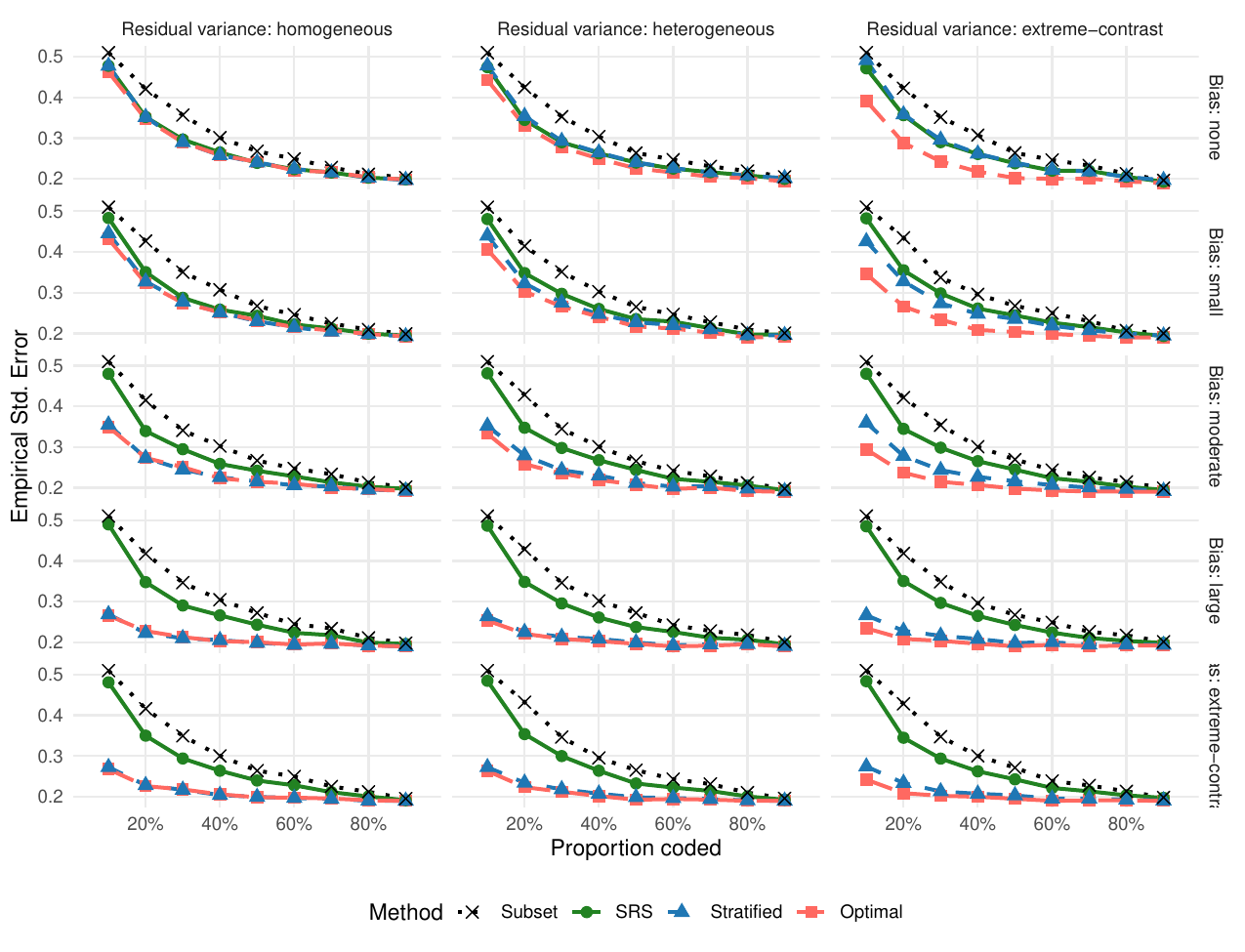}
    \caption{Empirical standard error of each estimator across simulation scenarios when the target $R^2 = 0.40$. }
    \label{fig:frontier_lowR2}
\end{figure}

\begin{figure}[h]
    \centering
    \includegraphics[width=0.9\textwidth]{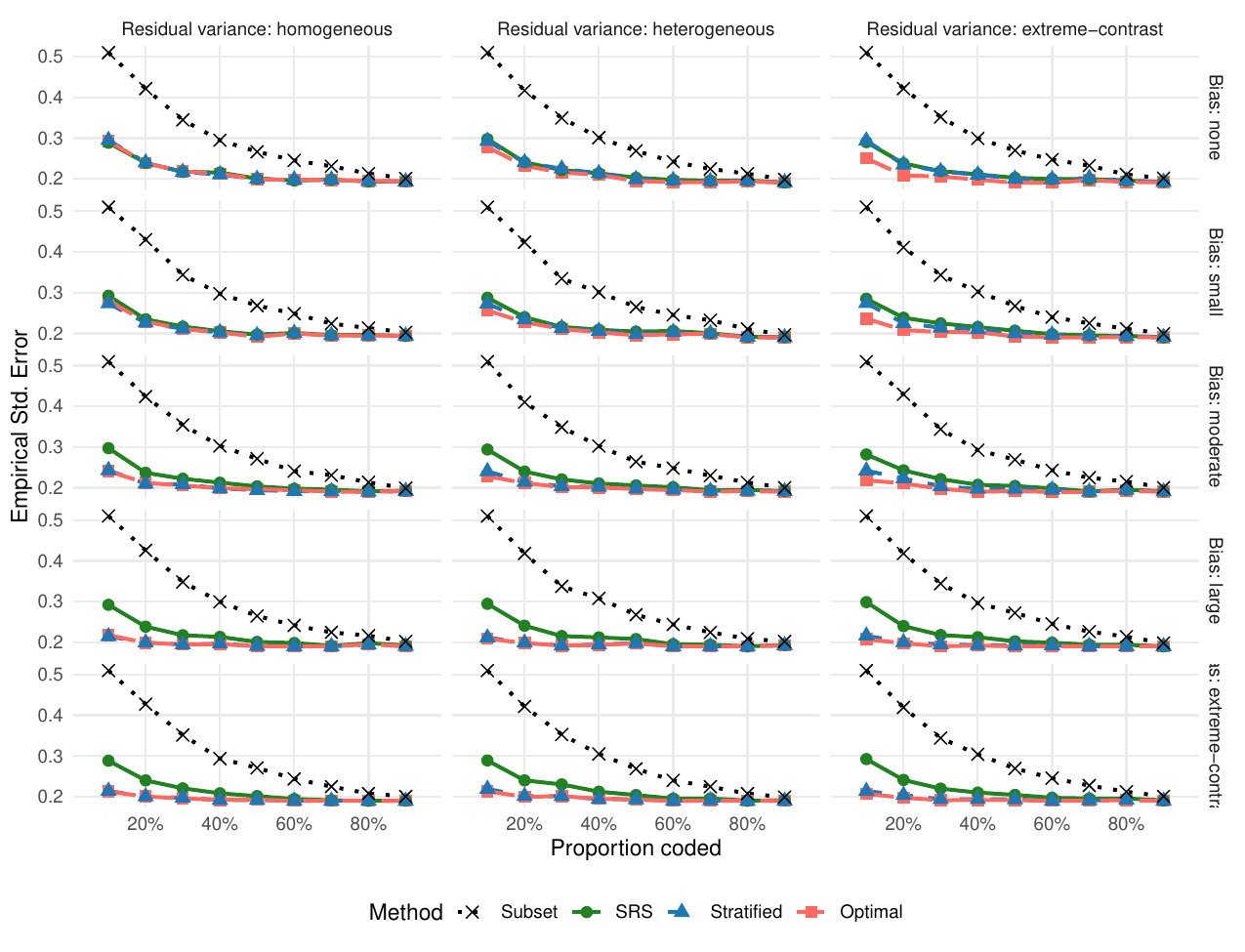}
    \caption{Empirical standard error of each estimator across simulation scenarios when the target $R^2 = 0.85$.}
    \label{fig:frontier_highR2}
\end{figure}

Figures~\ref{fig:frontier_lowR2} and \ref{fig:frontier_highR2} present the empirical standard errors of each estimator across the full grid of bias patterns (rows) and residual variance structures (columns), for target $R^2 = 0.40$ and $R^2 = 0.85$, respectively. 

Note: In all figures below, the subset estimator (black) uses only the human-coded sample; SRS (green) is the model-assisted estimator under simple random sampling; Stratified (blue) is the model-assisted estimator using proportional allocation, and Optimal (red) is the stratified model-assisted estimator using optimal (Neyman) allocation.

The disaggregated results yield several notable findings:

\begin{enumerate}
 \item \textbf{Model-assisted estimation universally dominates subset estimation.} Across all 15 configurations and both levels of $R^2$, the model-assisted estimators (SRS, Stratified, and Optimal) uniformly outperform the simple subset estimator. This confirms the theoretical results from \citet{mozer2024more} showing that incorporating surrogate predictions improves efficiency whenever $R^2 > 0$.

\item \textbf{Stratification benefits increase with magnitude of bias.} Comparing across rows within each figure, the gap between the SRS estimator (green) and the stratified estimators (blue and red) grows as bias increases from no bias to large/extreme-contrast biases. When there is no bias (top row), the three model-assisted estimators perform nearly identically. When bias is large or extreme (bottom rows), stratification provides further efficiency gains.

  \item \textbf{Optimal allocation provides incremental gains with heterogeneous variance.} Comparing across columns, the difference in performance between the stratified estimator using proportional allocation (blue) and optimal allocation (red) is most pronounced in the ``extreme-contrast'' variance setting (right column), particularly at low coding budgets. This reflects the benefit of oversampling high-variance strata when residual variance differs across strata.
  
  \item \textbf{Efficiency differences are amplified when surrogates are less predictive of human-coded outcomes.} Comparing Figures~\ref{fig:frontier_lowR2} and \ref{fig:frontier_highR2}, all estimators show higher standard errors when $R^2 = 0.40$ compared to $R^2 = 0.85$, as expected. However, the relative benefits of stratification over SRS are larger at lower $R^2$, likely because noisier surrogates allow for more residual variance to differ across strata.
\end{enumerate}

\subsection{Variance reduction from stratification}

\begin{figure}[h]
    \centering
    \includegraphics[width=0.8\textwidth]{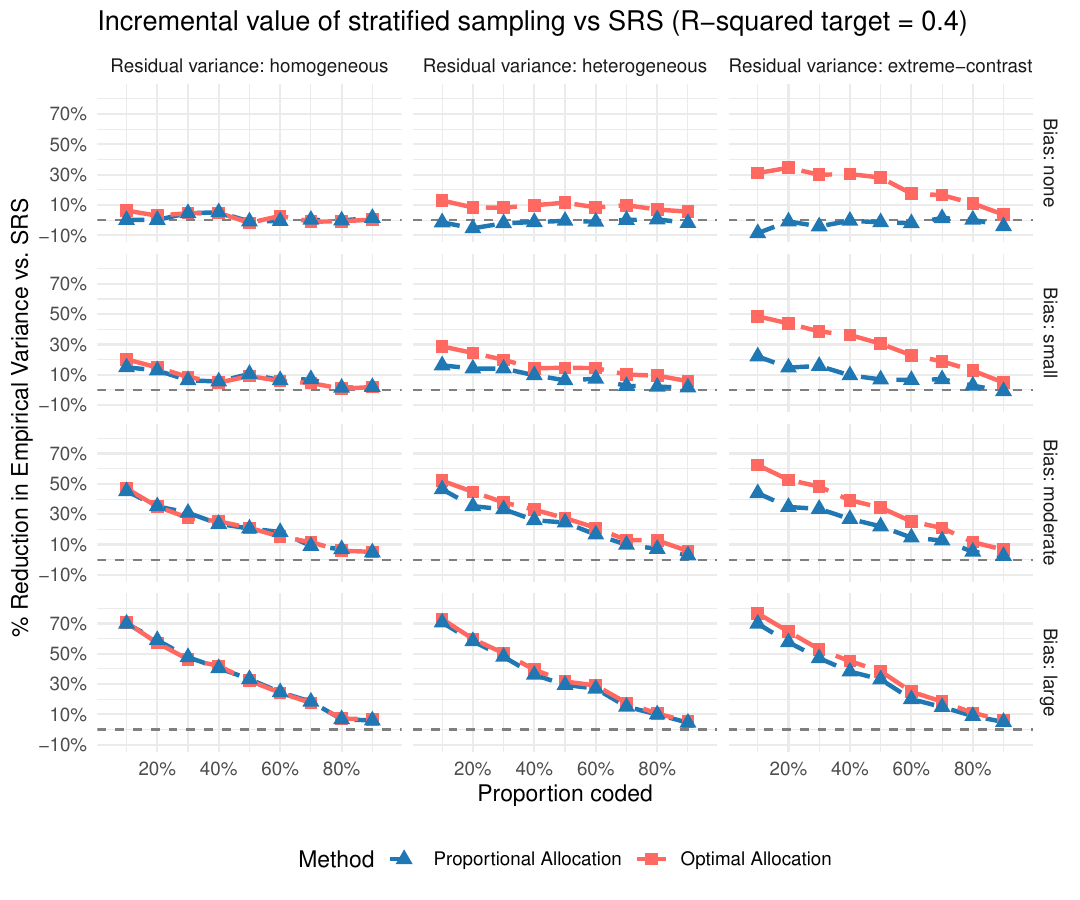}
    \caption{Percent reduction in empirical variance of the stratified model-assisted estimators relative to SRS when target $R^2 = 0.40$. Blue lines show proportional allocation; red lines show optimal (Neyman) allocation. The dashed horizontal line at 0\% indicates no improvement over SRS.}
    \label{fig:delta_lowR2}
\end{figure}

\begin{figure}[h]
    \centering
    \includegraphics[width=0.8\textwidth]{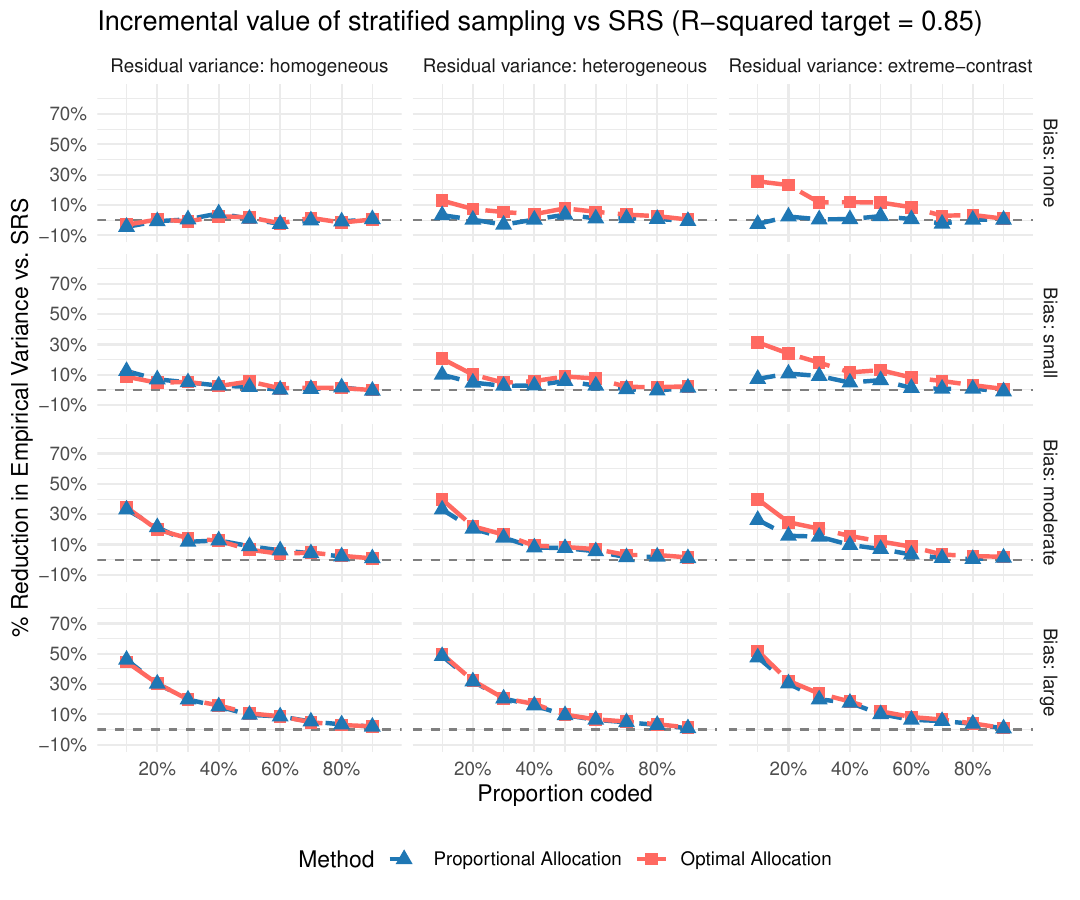}
    \caption{Percent reduction in empirical variance of the stratified model-assisted estimators relative to SRS when target $R^2 = 0.85$.}
    \label{fig:delta_highR2}
\end{figure}

Figures~\ref{fig:delta_lowR2} and \ref{fig:delta_highR2} show the percent reduction in empirical variance achieved by stratified sampling relative to the model-assisted estimator under SRS, broken down by all configurations of bias and residual variance. These figures extend Figure~\ref{fig:sims_re_vs_SRS} in the main text.

Again, several important patterns arise:

\begin{enumerate}
    \item \textbf{Structured residuals drive efficiency gains.} When there is no bias and residual variance is homogeneous (top-left panel of each figure), stratification provides essentially no benefit (variance reduction near 0\%), confirming that stratification only helps when surrogate outcomes have structured prediction errors.

    \item \textbf{Bias drives larger gains than heterogeneous variance.} Comparing across rows (varying bias) versus columns (varying residual variance), the largest variance reductions occur when bias increases (moving down the rows), with reductions exceeding 50--70\% at small coding budgets under the large and extreme-contrast bias settings. In comparison, comparing columns within the ``no bias'' row, we see that heterogeneous residual variance alone leads to more modest gains of 10--30\%.

    \item \textbf{Optimal allocation adds incremental value with heterogeneous variance.} The gap between the blue (proportional) and red (optimal) lines is largest in the rightmost column (extreme-contrast residual variance), where optimal allocation can provide an additional 5--15 percentage points of variance reduction beyond proportional allocation.
\end{enumerate}

\subsection{Variance inflation relative to full coding}

Figures~\ref{fig:vs_full_lowR2} and \ref{fig:vs_full_highR2} below show the variance inflation factor, defined as the ratio of each estimator's variance to the variance of the oracle (full-coding) estimator, across all simulation configurations. A variance ratio of 1 indicates performance equivalent to full coding; higher ratios indicate less efficient estimation.

\begin{figure}[h]
    \centering
    \includegraphics[width=0.9\textwidth]{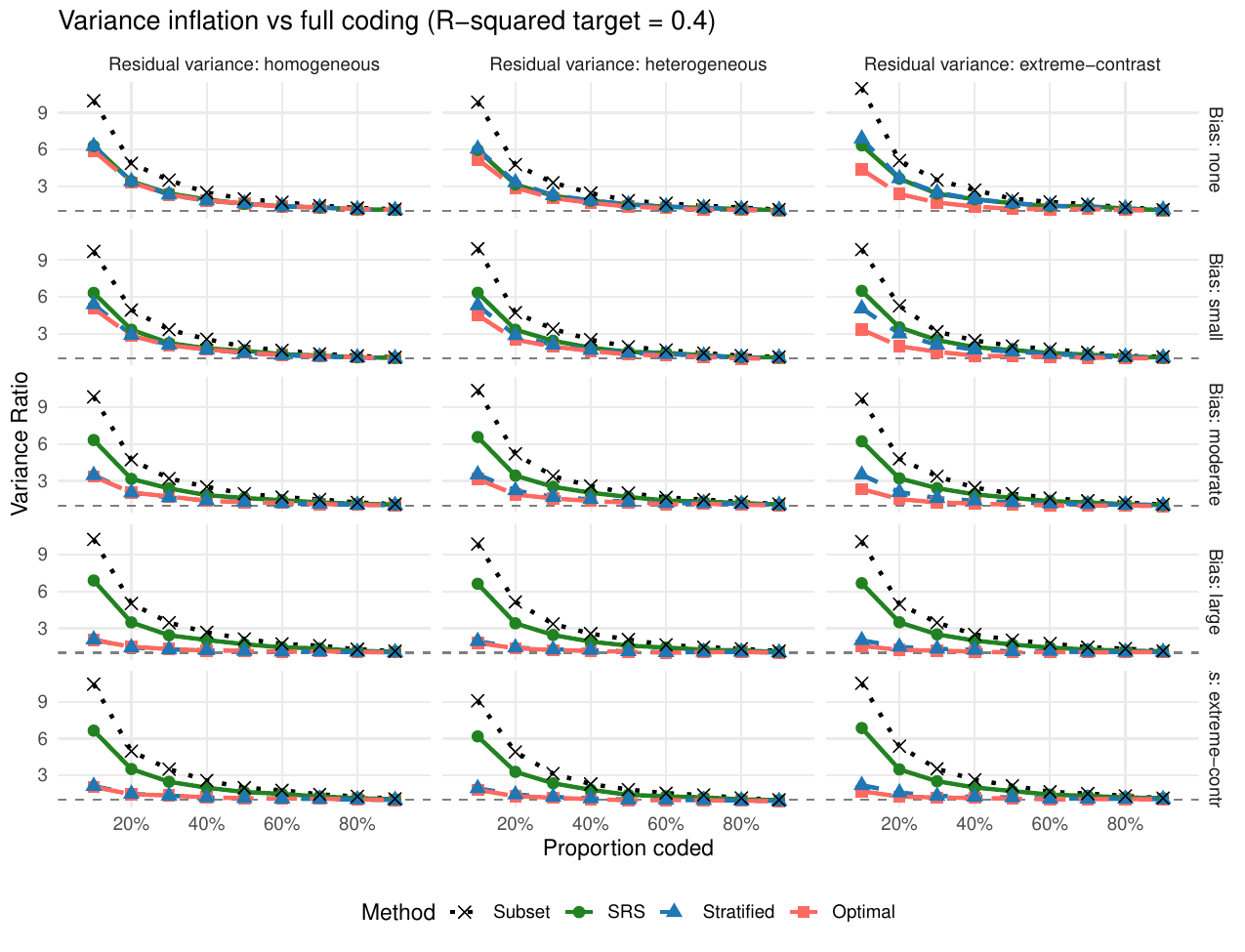}
    \caption{Variance inflation relative to full coding (oracle estimator) across simulation scenarios when target $R^2 = 0.40$.}
    \label{fig:vs_full_lowR2}
\end{figure}

\begin{figure}[h]
    \centering
    \includegraphics[width=0.9\textwidth]{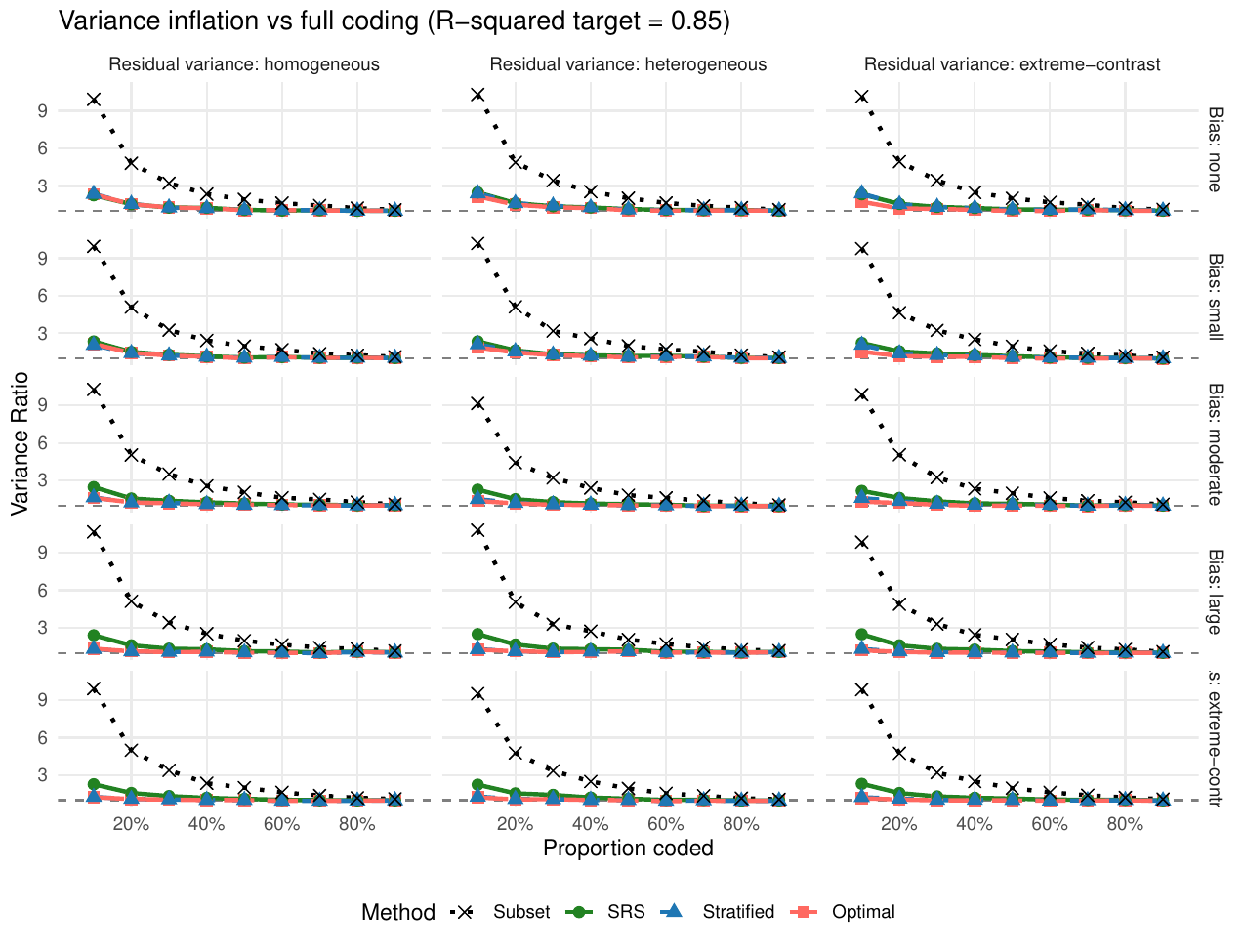}
    \caption{Variance inflation relative to full coding (oracle estimator) across simulation scenarios when target $R^2 = 0.85$.}
    \label{fig:vs_full_highR2}
\end{figure}


Key findings based on these results include:

\begin{enumerate}
    \item \textbf{High $R^2$ dramatically reduces variance inflation.} Comparing Figures~\ref{fig:vs_full_lowR2} and \ref{fig:vs_full_highR2}, the model-assisted estimators achieve variance ratios much closer to 1.0 when $R^2 = 0.85$ than when $R^2 = 0.40$. At $R^2 = 0.85$ with 30\% coding, the variance inflation is typically less than 2$\times$ for the stratified estimators, whereas at $R^2 = 0.40$, inflation can exceed 3--4$\times$.

    \item \textbf{Stratification approaches oracle efficiency in favorable conditions.} When bias is large and $R^2$ is high (bottom rows of Figure~\ref{fig:vs_full_highR2}), the stratified estimators achieve variance ratios very close to 1 even at relatively small coding budgets, demonstrating that strategic sampling can almost completely offset the loss due to incomplete coding.
\end{enumerate}

\subsection{Coverage and bias diagnostics}

To validate the theoretical properties of the estimators, here we present 95\% confidence interval coverage rates and Monte Carlo bias for each estimator across all simulation configurations. Table~\ref{tab:coverage_bias_summary} summarizes coverage and bias statistics aggregated across all scenarios.

\begin{table}[h]
\caption{Coverage and Monte Carlo bias by estimation method (aggregated across all simulation configurations)}
\label{tab:coverage_bias_summary}
\begin{tabular*}{\linewidth}{@{\extracolsep{\fill}}crrrrr}
\toprule
 & \multicolumn{3}{c}{95\% CI Coverage} & \multicolumn{2}{c}{Monte Carlo Bias} \\ 
\cmidrule(lr){2-4} \cmidrule(lr){5-6}
Method & Mean & Min & Max & Mean & Max $|$Bias$|$ \\ 
\midrule\addlinespace[2.5pt]
Subset & 95.3\% & 93.8\% & 96.4\% & -0.0004 & 0.0222 \\ 
SRS & 95.4\% & 94.2\% & 96.2\% & -0.0005 & 0.0143 \\ 
Stratified & 95.4\% & 94.2\% & 96.3\% & -0.0004 & 0.0143 \\ 
Optimal & 95.4\% & 94.1\% & 96.5\% & -0.0003 & 0.0120 \\ 
\bottomrule
\end{tabular*}
\end{table}

All four estimators achieve mean coverage rates very close to the nominal 95\% level, ranging from 95.3\% (Subset) to 95.4\% (SRS, Stratified, and Optimal).  These results confirm that the variance estimators derived in Section~\ref{sec:methods} yield valid confidence intervals across a wide range of conditions. Similarly, Monte Carlo bias is negligible for all estimators, with mean bias effectively zero (less than 0.001 in absolute value) and maximum absolute bias across all configurations below 0.025. This confirms the design-based unbiasedness property: the estimators are unbiased regardless of whether the surrogate model is correctly specified or exhibits systematic prediction errors across strata.

\section{Details for the empirical applications}
\subsection{Generating LLM predictions as surrogate outcomes}
\label{app:chatgpt}

We used OpenAI's \texttt{gpt-4o-mini} model to generate predicted writing quality scores for both empirical applications presented in Section~\ref{sec:application}. The LLM predictions were generated using the \texttt{ellmer} R package \citep{ellmer}, which provides an interface for extracting structured outputs from ChatGPT through the OpenAI API \citep{openai_api}. Below we describe the prompting strategy for each application.

\subsubsection{MORE Study (Grades 1--2 Essays)}

For the MORE study, essays were written by students in grades 1--2 in response to four different prompts covering science and social studies topics. Each essay was scored by human coders using a 0--8 point rubric evaluating four dimensions: Claim (0--2 points), Evidence (0--4 points), Ending (0--1 point), and Clarity/Voice (0--1 point). The holistic score was computed as the sum of these dimension scores. See \citet{kim2021improving2} for a complete description of the experiment and human coding process.

To generate surrogate outcomes for these data, we used the following system prompt:

\begin{quote}
\textit{You are an expert essay grader for students in grades 1--2. Given an essay text, your job is to evaluate and score the overall quality of the essay. Use a scale from 0 to 8, where higher score indicates greater quality. 
Your evaluation should consider the following criteria: (1) Claim (0--2 points), (2) Evidence (0--4 points), (3) Ending (0--1 points), (4) Clarity/Voice (0--1 points).}

\textit{When evaluating an essay, first code for the 4 dimensions above. Then derive an overall quality score as a function of the dimension scores. Note that the term `XXX' in an essay indicates a word that was not legible or could not be transcribed from the handwritten essays. Present your response as a numerical value.}   

\textit{In addition, provide a confidence rating for each essay indicating how confident you are in your quality score on a scale from 0-100, where:}
\begin{itemize}
\tightlist
	 \item \textit{86-100: Very confident, clear-cut case}
   \item \textit{70-85: Fairly confident, minor ambiguity}
   \item \textit{50-69: Moderate confidence, could reasonably be adjacent score}
   \item \textit{Below 50: Low confidence, difficult to assess}
   
\end{itemize}

   \textit{Provide only the overall score and confidence rating for each essay, without any explanation or additional text.}
\end{quote}

Each individual essay was then submitted to the LLM with the user prompt:

\begin{quote}
\textit{Please score the following essay, which was written in response to the prompt ``[PROMPT]'': [ESSAY TEXT]}
\end{quote}

\noindent where [PROMPT] was the writing prompt assigned to the student and [ESSAY TEXT] was the student's essay.

\subsubsection{PERSUADE Corpus (Grades 6--12 Essays)}

For the PERSUADE corpus, essays were written by students in grades 6--12 in response to one of fifteen different prompts (including 8 ``independent" prompts where students were asked to develop arguments based on their own reasoning and 7 ``Text dependent'' prompts where students were asked to draw on provided source materials). All essays were scored by trained human coders using a holistic 1--6 point rubric. See \citet{persuade2} for a complete description of the data and human scoring process.

To generate surrogate outcomes, the LLM was prompted with the following system prompt:

\begin{quote}
\textit{You are an expert essay grader for students in grades 6-12. Given an essay text, your job is to evaluate and score the overall quality of the essay.
Use a scale from 1 to 6 , where higher score indicates greater quality. Present your response as a numerical value. }
 
 \textit{In addition, provide a confidence rating for each essay indicating how confident you are in your quality score on a scale from 0-100, where:}
\begin{itemize}
\tightlist
	 \item \textit{86-100: Very confident, clear-cut case}
   \item \textit{70-85: Fairly confident, minor ambiguity}
   \item \textit{50-69: Moderate confidence, could reasonably be adjacent score}
   \item \textit{Below 50: Low confidence, difficult to assess}
   
\end{itemize}

   \textit{Provide only the overall score and confidence rating for each essay, without any explanation or additional text.}
\end{quote}

Each individual essay was then submitted with the user prompt:

\begin{quote}
\textit{Please score the following essay, which was written by a student in grade [GRADE] in response to the prompt ``[PROMPT]'': [ESSAY TEXT]}
\end{quote}

\noindent where [GRADE] was the student's grade level, [PROMPT] was the assigned writing prompt, and [ESSAY TEXT] was the student's essay. 

\subsubsection{Technical Implementation}

All LLM predictions were generated using OpenAI's \texttt{gpt-4o-mini} model via the \texttt{ellmer} R package \citep{ellmer}. We used structured outputs (\texttt{type\_object()} in \texttt{ellmer}) to ensure the LLM returned properly formatted numeric responses. Requests were processed in parallel using \texttt{parallel\_chat\_structured()} with a maximum of 200 concurrent requests and a rate limit of 1000--2000 requests per minute to comply with OpenAI's API rate limits \citep{openai_api}.

The total cost for generating all predictions was approximately \$0.42 for the MORE study ($N = 5294$ essays) and \$4.51 for the PERSUADE corpus ($N = 25996$ essays).

\subsection{Comparison of stratification techniques}
\label{app:strat_techniques}

To implement the proposed stratified model-assisted estimation approach in practice, researchers must first decide how to stratify their data before any human coding has occurred. This presents a challenge: the ideal stratification would be based on the residual variance structure $\sigma^2_{zk} = \text{Var}(Y_i - \hat{Y}_i \mid Z_i = z, K_i = k)$, but these residuals cannot be observed prior to coding. To address this issue, we developed a practical approach for constructing strata using only aspects of the data that are available prior to coding (e.g., the surrogate outcomes $\hat{Y}$ and other machine-extractable summary measures of the data.

\subsubsection{Candidate Stratification Variables}

For our empirical applications, we considered stratifications based on three types of variables that can be measured before human coding: (1) the LLM-predicted quality scores ($\hat{Y}$), (2) the LLM's self-reported confidence ratings, and (3) simple summary measures of the text such as word count. For each variable, we first created single-variable stratifications using quantile-based cutoffs (tertiles, quartiles, and quintiles).\footnote{For the MORE study, quantile cutoffs were computed separately within each treatment arm to ensure balanced stratum sizes across arms.} We also constructed stratifications based on interactions between variables by crossing pairs of variables at different granularities (2$\times$2, 2$\times$3, 3$\times$2, and 3$\times$3 configurations). This resulted in approximately 20--25 candidate stratifications per application.

The \texttt{create\_stratification\_candidates()} function in the \texttt{stratsampling} package implements this strategy, taking a data frame and vector of variable names as inputs and returning a list of candidate stratifications based on those variables.

\subsubsection{Pre-Coding Evaluation Criteria}

Since residuals are unavailable before coding, we compared candidate stratifications using proxy metrics computed solely from the surrogate outcomes $\hat{Y}$. Specifically, we considered: 
\begin{itemize}
    \item \textbf{Variance of stratum means} ($\text{Var}_k[\bar{\hat{Y}}_k]$): Higher values indicate greater separation of strata, which often correlates with systematic differences in prediction bias across strata. This serves as a proxy for the between-strata variance (BS) component that drives efficiency gains from stratification.

    \item \textbf{Stratum size balance}: The ratio of the largest to smallest stratum size. Highly unbalanced strata may reduce efficiency.

    \item \textbf{Minimum stratum size}: Small within-stratum sample sizes may lead to unstable variance estimates.
\end{itemize}

The \texttt{compare\_stratifications\_precoding()} function in the \texttt{stratsampling} package implements this evaluation, ranking candidate stratifications by the variance of stratum means.

In both of our empirical applications, we computed the above metrics for each candidate stratification selected the stratification with the largest variance of stratum means. Candidates with highly unbalanced stratum sizes (ratio $> 10$) or very small strata ($N_{zk} < 100$) were excluded from consideration to ensure reliable variance estimation.

\subsubsection{Final Stratification Choices}

\paragraph{MORE Study.} Based on the pre-coding evaluation, our selected stratification was defined by quantiles of the LLM-predicted quality scores, yielding $K$=4 strata per treatment arm. Table~\ref{tab:reads_strat_v2} in the main text shows the characteristics of these strata. 

\paragraph{PERSUADE Corpus.} For the PERSUADE corpus, our selected stratification was defined by the interaction  between LLM-predicted quality\footnote{For the PERSUADE corpus, LLM predictions were aggregated across the lowest scores ($\hat{Y}\in\{1,2\}$) and the highest scores ($\hat{Y}\in\{5,6\}$) when generating candidate stratifcations, as only a small number of essays had predictions of 1 or 6.} (1-2, 3, 4, 5-6) and essay length (word count; below or above the median of 375 words). Table~\ref{tab:persuade_strat} in the main text shows the characteristics of these strata.

\subsubsection{Oracle Evaluation (Retrospective Validation)}

For validation purposes, we also computed oracle metrics for each candidate stratification using the actual residuals (available only because all data was already coded in both of our applications). Examining these metrics (e.g., the ratio of maximum to minimum within-stratum residual variance and the between-strata variance of mean residuals) allowed us to verify that our pre-coding proxy metrics identify stratifications that would perform well in practice. The \texttt{compare\_stratifications()}  and \texttt{compare\_decomposition\_oracle()} functions in the \texttt{stratsampling} package implement this evaluation.

\paragraph{Variance Reduction Across Candidate Stratifications.}
Using the oracle residuals $e_i = Y_i - \hat{Y}_i$, we computed the variance decomposition from Theorem~\ref{theorm:var_reduction} for each candidate stratification. Specifically, we calculated the between-strata (BS) and within-strata (WS) components of variance, where the difference $\text{BS} - \text{WS}$ represents the theoretical variance reduction achieved by stratified sampling relative to SRS under proportional allocation. Positive values indicate that stratification reduces variance compared to SRS.

\begin{figure}[h]
    \centering
    \includegraphics[width=0.85\textwidth]{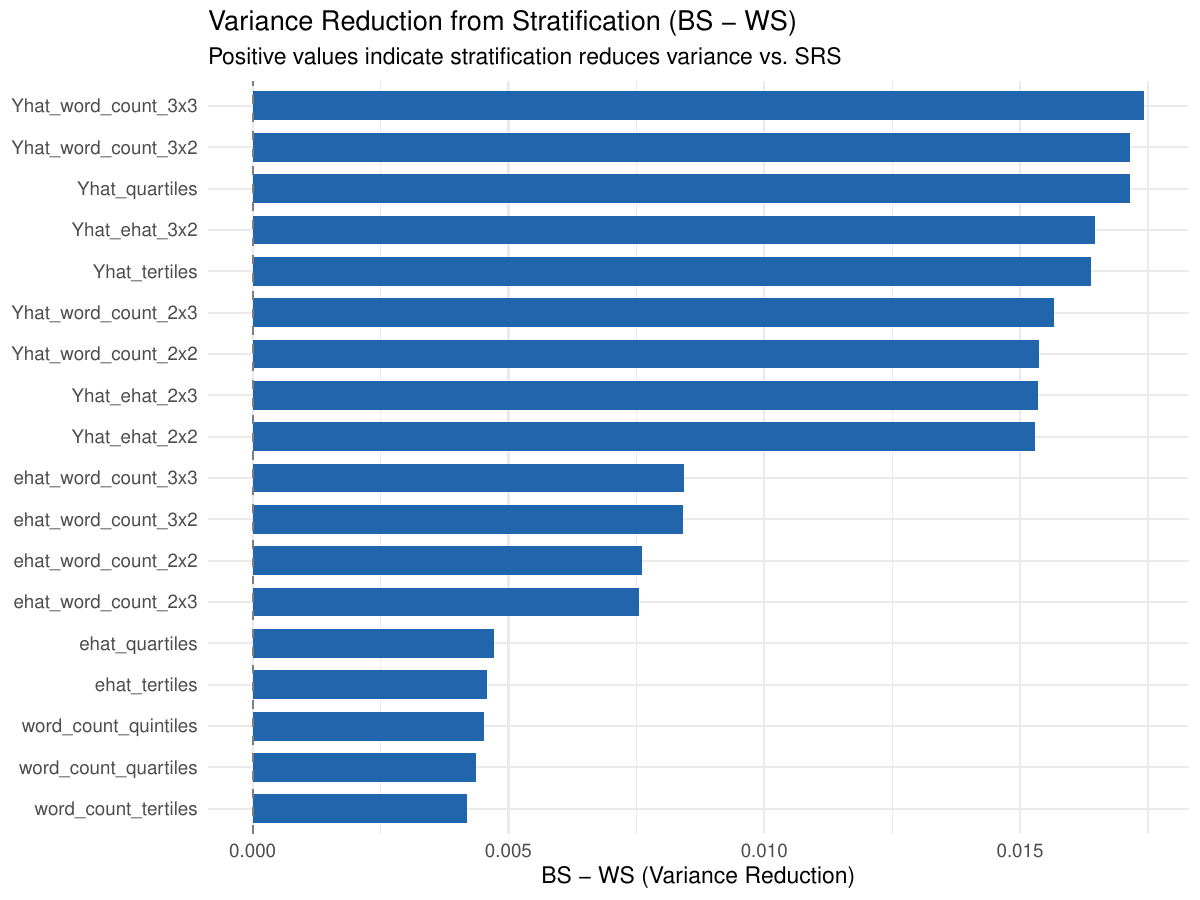}
    \caption{Oracle variance reduction ($\text{BS} - \text{WS}$) from stratification for each candidate stratification in the MORE study.}
    \label{fig:reads_var_reduction}
\end{figure}

\begin{figure}[h]
    \centering
    \includegraphics[width=0.85\textwidth]{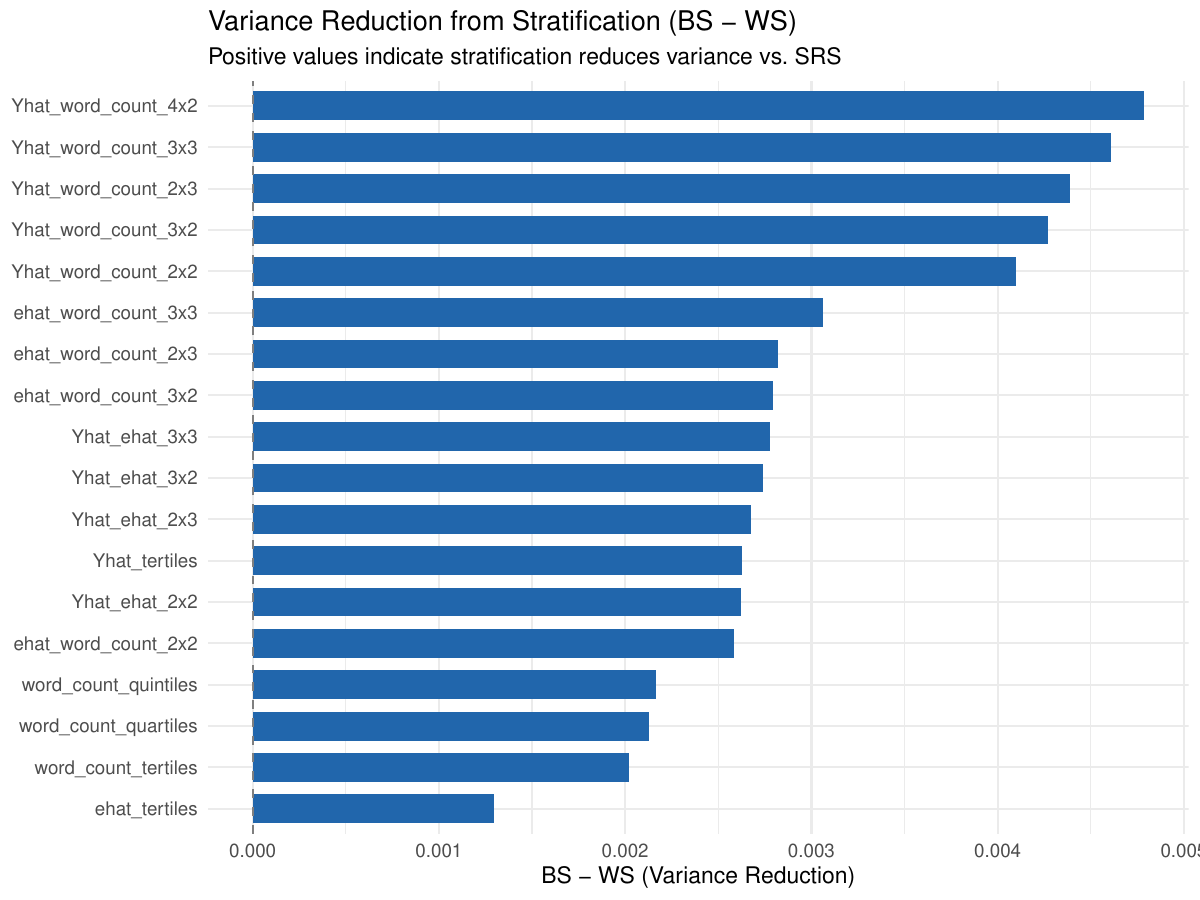}
    \caption{Oracle variance reduction ($\text{BS} - \text{WS}$) from stratification for each candidate stratification in the PERSUADE corpus.}
    \label{fig:persuade_var_reduction}
\end{figure}

Figures~\ref{fig:reads_var_reduction} and ~\ref{fig:persuade_var_reduction} show the calculated values of the oracle variance reduction ($\text{BS} - \text{WS}$) for each candidate stratification in the MORE study and PERSUADE applications, respectively. 

In both applications, all candidate stratifications yield positive variance reductions, confirming that stratification improves efficiency relative to SRS regardless of the specific stratification variable chosen. However, the magnitude of improvement varies substantially across candidates. Stratifications based on the LLM predictions ($\hat{Y}$) generally outperform those based on word count alone, and combining $\hat{Y}$ with word count in a crossed design (e.g., $3 \times 3$) tends to achieve the largest variance reductions. Single-variable stratifications based on word count or residual predictions ($\hat{e}$) alone provide more modest gains.

\paragraph{Validation of Pre-Coding Proxy Metrics.}
A key practical question is whether the pre-coding proxy metrics (computed without access to oracle residuals) successfully identify the stratifications that achieve the best oracle performance. To assess this, we compared the pre-coding proxy calculated as the variance of stratum means of the surrogate predictions, $\text{Var}_k[\bar{\hat{Y}}_k]$ against the oracle metric of average within-stratum residual variance, $\text{Var}_k[\bar{e}_k]$, across all candidate stratifications.

\begin{figure}[h]
    \centering
    \includegraphics[width=0.75\textwidth]{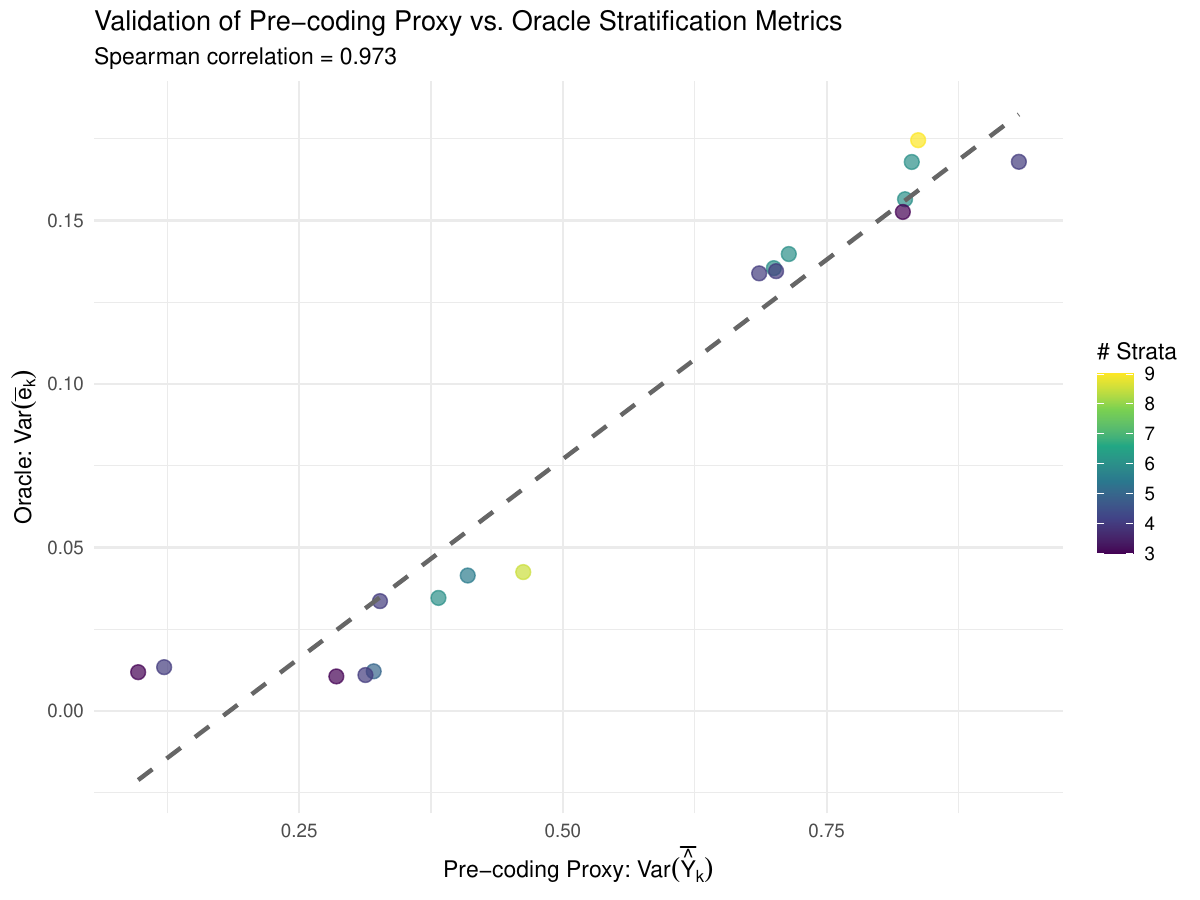}
    \caption{Validation of pre-coding proxy metrics for comparing candidate stratifications in the MORE study.}
    \label{fig:reads_validation}
\end{figure}

\begin{figure}[h]
    \centering
    \includegraphics[width=0.75\textwidth]{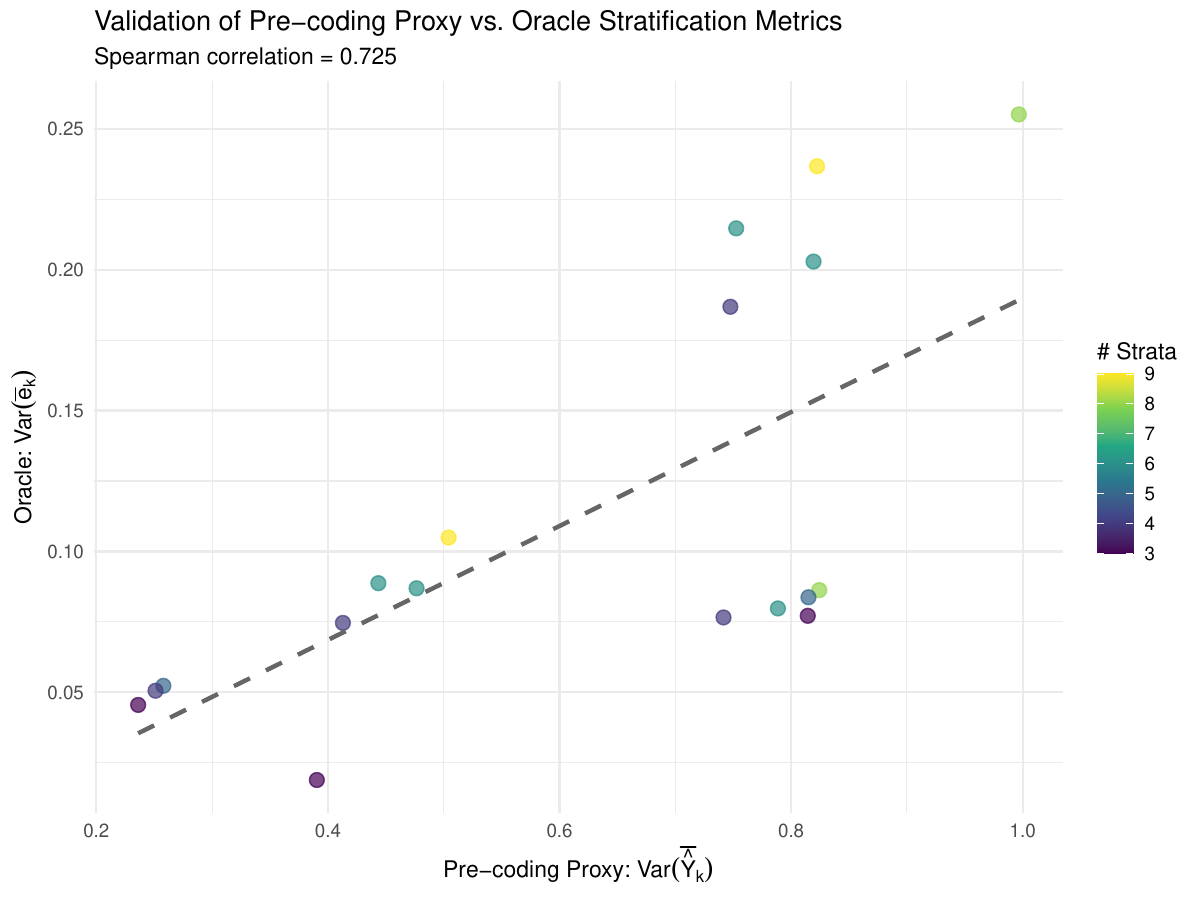}
    \caption{Validation of pre-coding proxy metrics for comparing candidate stratifications in the PERSUADE corpus.}
    \label{fig:persuade_validation}
\end{figure}

Figures~\ref{fig:reads_validation} and ~\ref{fig:persuade_validation} display these comparisons for the MORE study and PERSUADE applications, respectively. Each point represents one candidate stratification, with the x-axis showing the pre-coding proxy and the y-axis showing the oracle metric. Points are colored by the number of strata in each stratification.

The results show that the pre-coding proxy metric is a reliable indicator of oracle performance. In the MORE study (Figure~\ref{fig:reads_validation}), the Spearman correlation between the pre-coding proxy and oracle metric is 0.973, indicating near-perfect agreement in the ranking of candidate stratifications.  In the PERSUADE corpus (Figure~\ref{fig:persuade_validation}), the Spearman correlation is somewhat lower at 0.725, reflecting greater heterogeneity in the relationship between surrogate predictions and oracle residuals in this more complex dataset. Nevertheless, the positive correlation indicates that the pre-coding proxy still provides useful guidance: stratifications with higher variance of stratum means in the surrogate predictions tend to achieve better oracle performance, even if the relationship is not perfectly monotonic. Importantly, the top-ranked stratifications by the pre-coding proxy (those with highest $\text{Var}_k[\bar{\hat{Y}}_k]$) consistently appear among the better performers according to the oracle metric.

These validation results support the practical utility of our proposed approach: researchers can use the pre-coding proxy metrics to select stratifications before any human coding occurs, with reasonable confidence that the selected stratification will achieve good efficiency gains in practice.

\subsection{Sensitivity to choice of coded sample}
\label{app:empirical_sens}

A natural concern when presenting results from a single iteration of stratified sampling (e.g., Tables~3 and~5 of the main text) is that the particular sample selected for human coding was ``cherry-picked'' to produce favorable results. To address this concern, we conducted a repeated sampling exercise that demonstrates the consistency of our estimators across different random selections of the coded sample.

For each empirical application, we repeated the following procedure 20 times:
\begin{enumerate}
    \item Draw a new random sample of 30\% of documents for human coding using each sampling method (SRS, proportional stratified, and optimal stratified allocation)
    \item Apply the corresponding estimator to obtain a point estimate
    \item Record the estimated treatment effect (MORE) or population mean (PERSUADE)
\end{enumerate}

This exercise allows us to visualize the sampling distribution of each estimator and assess whether the single-iteration results presented in the main text are representative of typical performance.

\subsubsection{MORE Study: Estimating a treatment effect}

\begin{figure}[h]
    \centering
    \includegraphics[width=0.85\textwidth]{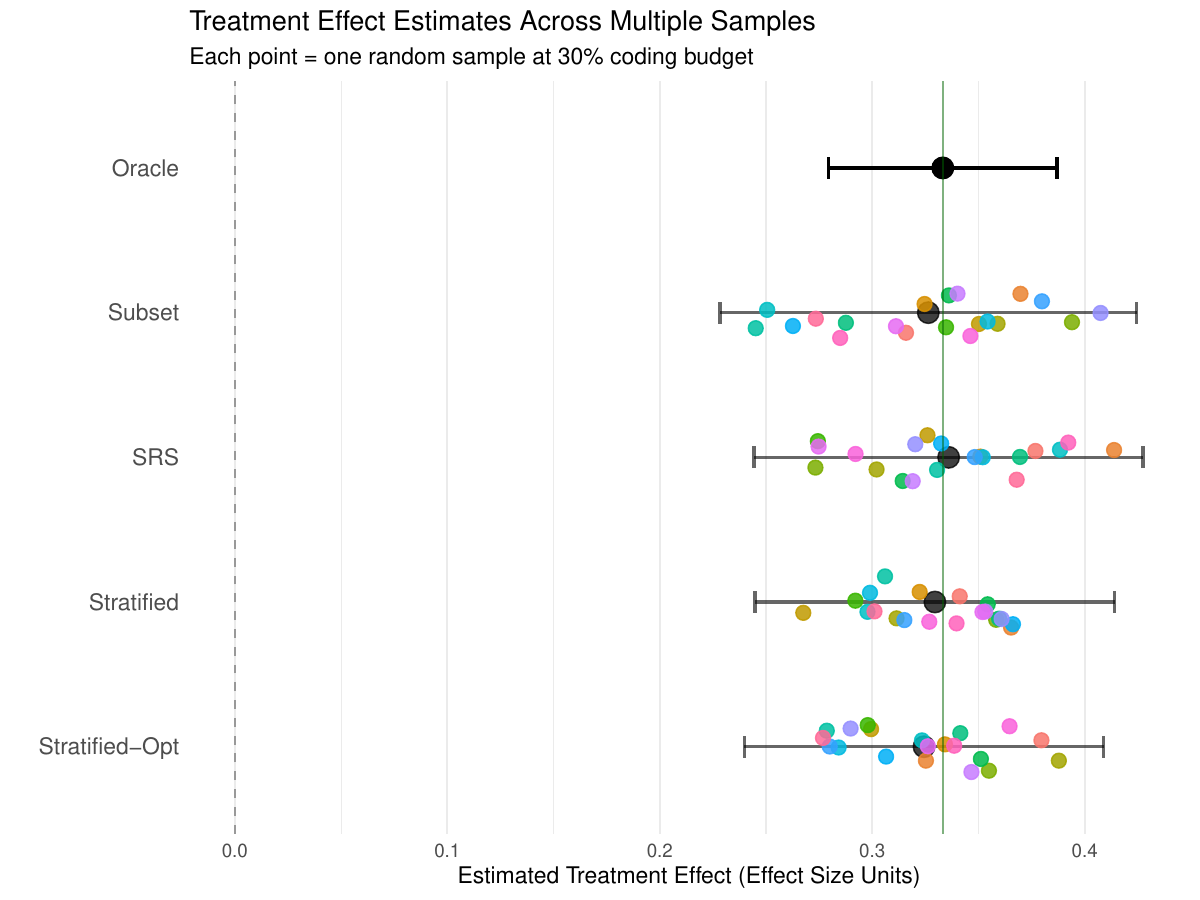}
    \caption{Treatment effect estimates for the MORE study across 20 repeated samples at $h = 0.30$ coding budget.}
    \label{fig:more-sensitivity}
\end{figure}

Figure~\ref{fig:more-sensitivity} displays the treatment effect estimates obtained across 20 independent samples for each estimation method in the MORE study application. Here, each colored point represents the estimate from one random sample; the black point shows the mean estimate across samples, with error bars representing $\pm 1.96$ times the average estimated standard error. The vertical green line indicates the oracle (full-coding) estimate. The horizontal spread of points within each method illustrates the sampling variability inherent in selecting different 30\% subsets for human coding.

Here we see that all methods consistently identify a positive and statistically significant treatment effect, with point estimates clustering around the oracle value of 0.333. The single-iteration results reported in Table~\ref{tab:reads_point_ests_v2} of the main text fall well within the range of estimates observed across repeated samples, confirming that those results were not anomalous.

\subsubsection{PERSUADE Corpus: Estimating a single mean}

\begin{figure}[h]
    \centering
    \includegraphics[width=0.85\textwidth]{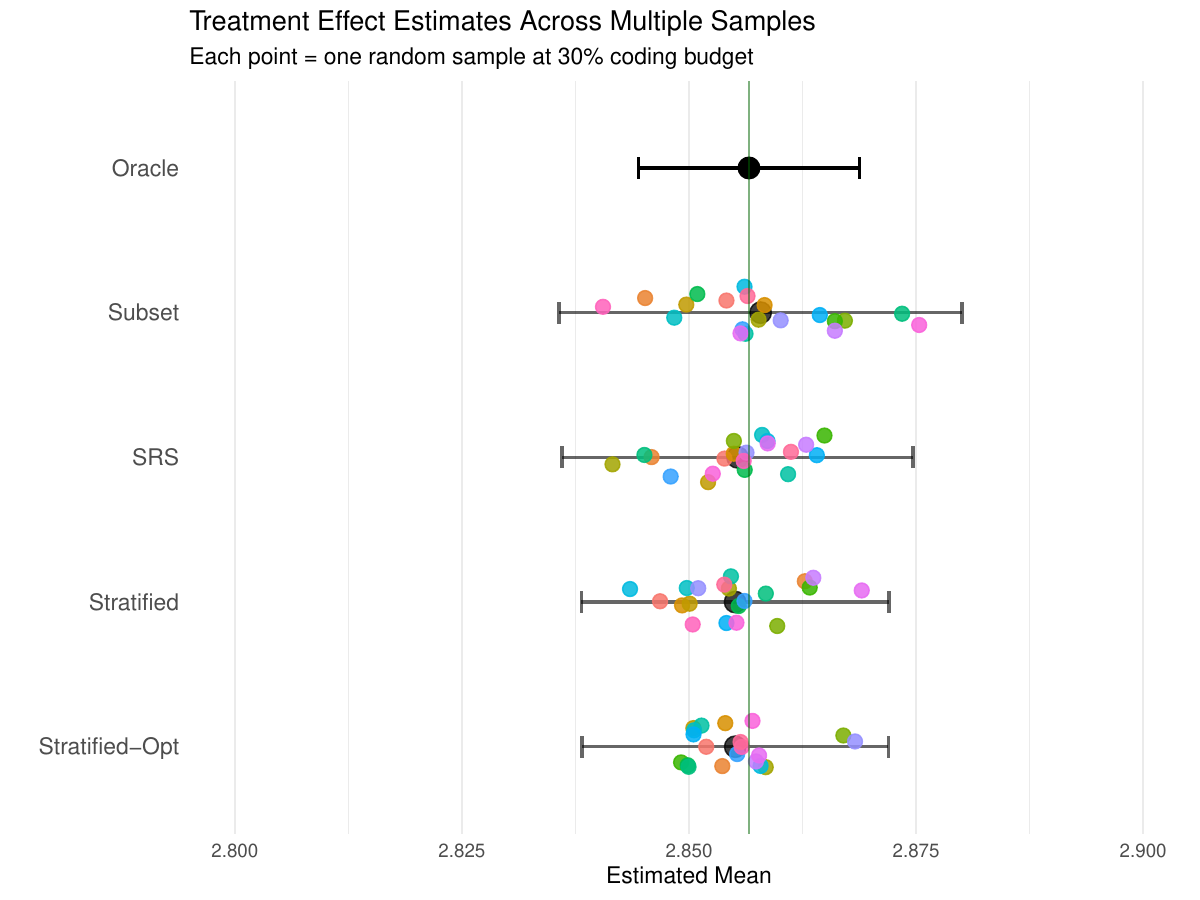}
    \caption{Estimated population mean for the PERSUADE corpus across 20 repeated samples at $h = 0.30$ coding budget.}
    \label{fig:persuade-sensitivity}
\end{figure}

Figure~\ref{fig:persuade-sensitivity} presents analogous results for the PERSUADE corpus application, where the goal is to estimate the population mean writing quality score.

The patterns observed are consistent with those from the MORE study. All methods produce estimates tightly clustered around the oracle value of 2.857, with the single-iteration results from Table~\ref{tab:persuade_point_ests} of the main text falling squarely within the observed range across repeated samples. The model-assisted estimators again show reduced variability compared to the subset  estimator, and the stratified approaches demonstrate the tightest clustering (i.e., less variability in point estimates).

\end{appendices}


\end{document}